\documentclass[12pt]{spieman}  
\usepackage{amsmath,amsfonts,amssymb}
\usepackage{graphicx}
\usepackage{setspace}
\usepackage{tocloft}
\usepackage{booktabs}
\usepackage{subcaption}
\usepackage{lineno}
\usepackage{tikz}
\usepackage[mode=buildnew]{standalone}

\DeclareMathOperator*{\argmin}{argmin}
\DeclareMathOperator*{\argmax}{argmax}

\title{Level set image segmentation with velocity term learned from data with applications to lung nodule segmentation}

\author[a]{Matthew C Hancock}
\author[a]{Jerry F Magnan}
\affil[a]{Florida State University, Department of Mathematics, 208 Love Building, 1017 Academic Way, Tallahassee, FL 32306, USA}

\newcommand{\R}{\mathbb{R}}
\cftpagenumbersoff{figure}
\cftpagenumbersoff{table}

\begin{document}

\maketitle


\begin{abstract}

\vspace{2ex}\noindent\textbf{Purpose}: Lung nodule segmentation, i.e., the algorithmic delineation of the lung nodule surface, is a fundamental component of computational nodule analysis pipelines. We propose a new method for segmentation that is a machine learning-based extension of current approaches, using labeled image examples to improve its accuracy.

\vspace{2ex}\noindent\textbf{Approach}: We introduce an extension of the standard level set image segmentation method where the velocity function is learned from data via machine learning regression methods, rather than a priori designed. Instead, the method employs a set of features to learn a velocity function that guides the level set evolution from initialization.

\vspace{2ex}\noindent\textbf{Results}: We apply the method to image volumes of lung nodules from CT scans in the publicly available LIDC dataset, obtaining an average intersection over union score of 0.7185~($\pm$0.1114), which is competitive with other methods. We analyze segmentation performance by anatomical location and appearance categories of the nodules, finding that the method performs better for isolated nodules with well-defined margins. We find that the segmentation performance for nodules in more complex surroundings and having more complex CT appearance is improved with the addition of combined global-local features.

\vspace{2ex}\noindent\textbf{Conclusions}: The level set machine learning segmentation approach proposed herein is competitive with current methods. It provides accurate lung nodule segmentation results in a variety of anatomical contexts.
\end{abstract}


{\noindent\footnotesize\textbf{*}Matthew C Hancock \linkable{mhancock743@gmail.com}}

\begin{spacing}{1}   

\section{Introduction}
\label{section:intro}
Lung cancer has the highest mortality rate of all cancers in both males and females in the United States~\cite{siegel_cancer_2017}. A lung nodule is a small to medium sized (roughly, 3~mm to 30~mm) abnormal region with a somewhat well-defined boundary. The definition is inherently imprecise because it is based on visual examples and the subjective interpretations thereof~\cite{mcnitt-gray_lung_2007}. The likelihood of malignancy of a lung nodule can often be inferred by a combination of radiological features (e.g., growth rate, shape, or density features)~\cite{erasmus_solitary_2000} that if determined early increases the chance of survival~\cite{henschke_survival_2006}. The precise location of the nodule's surface in the image volume is necessary to produce such nodule features computationally. Thus, the accurate segmentation of the nodule is a crucial step in a computer-aided diagnosis (CAD) system. However, lung nodule segmentation is challenging because of the variability in nodule appearance and shape, as well as the potential proximity to structures in the lung anatomy (e.g., to the vasculature and pleural wall).

In this work, we introduce a machine learning extension of the standard level set image segmentation method of Malladi and Sethian~\cite{malladi_shape_1995} and apply it to the lung nodule segmentation problem. Starting from an initial guess of the boundary, the method evolves a function that moves toward the true boundary of the lung nodule contained in the image volume. The evolution of this function is governed by a partial differential equation (PDE) whose velocity term is a function of the underlying image data, including its local and global features. The standard level set approach requires the manual and a priori design of this velocity term, which is difficult since it requires prescribing precisely how the velocity function should depend on the underlying image data in all anticipated contexts. In our extension of this method, which we call the level set machine learning (LSML) method, the velocity term in the PDE is learned from data via machine learning regression models. This method makes successive estimates of the velocity function through a set of extracted local, global, image, and shape features. The relationship between the chosen features and the target velocity function that generates the motion of the evolving segmentation is approximated via regression.

This paper is structured as follows: in Section~\ref{section:background} we provide background on the level set segmentation method, and on related works and their results for lung nodule segmentation. In Section~\ref{section:methods}, we introduce the LSML image segmentation method, and in Section~\ref{section:results}, we present and discuss the results from applying our method to the lung nodule segmentation problem. We conclude in Section~\ref{section:conclusions}.

\section{Background}
\label{section:background}
\subsection{Level set image segmentation and variants}
Segmentation of images by evolving contours was introduced by Kass et al.~\cite{kass_snakes:_1988} where a parameterized curve is evolved by minimizing a weighted sum of internal and external energy functionals. Evolving a parametrized curve is tedious from a computational point of view because topological changes are not easily handled. On the other hand, implicit curves handle topological changes very naturally. In the level set method, the curve denoting an object's boundary is given implicitly by the zero level curve of a function $u$, which is often called the level set function. This concept extends readily to higher dimensions where the zero level set is a surface. The level set approach was pioneered by Osher and Sethian \cite{osher_fronts_1988} for tracking flame movement, and Malladi and Sethian~\cite{malladi_shape_1995} introduced the level set method into the realm of image segmentation. 

The movement of the level set function $u$ is governed by the following PDE\cite{malladi_shape_1995},
\begin{equation}
\label{eq:lspde}
    u_t = \nu \|Du\|
\end{equation}
where the term $\nu$ governs the velocity in directions orthogonal to the level sets of $u$. To perform image segmentation, the velocity $\nu$ in Eq.~\eqref{eq:lspde} is defined in terms of the underlying image information. Intuitively, the velocity of a point on the evolving segmentation boundary should be positive in the interior of the target boundary to cause expansion, and negative outside of the target boundary to cause contraction. Near the target boundary, the velocity should be small in magnitude to prevent overshooting it. The manual design of such a velocity field is difficult and often entails making simplifying assumptions, such as that of a homogeneously bright object against a homogeneously dark background. Such assumptions are often violated in practice, since the object to be segmented and its boundaries can be fuzzy, partially occluded, have heterogeneous brightness, and generally violate the assumptions used for modeling in numerous unpredictable ways. These difficulties often occur in medical imagery, where the lung nodules are attached, or are in close proximity, to separate anatomical objects with locally similar appearance, e.g., as in juxta-pleural and juxta-vascular lung nodules. The result of making simplistic or faulty assumptions is that the zero level set fails to evolve towards the desired true nodule boundary, thus excluding portions of the nodule or including non-nodule regions as part of a nodule.

In typical level set image segmentation approaches, statistical information collected from a dataset of training examples is not used. A few works have explored this avenue, mostly by attempting to enforce the evolving function $u$ to conform to more statistically likely shapes, or to enclose image regions that are statistically more similar in appearance. The first work to introduce such an approach is due to Leventon~\cite{leventon2002statistical}, where principal component analysis (PCA) (and the distribution of its coefficients) was applied to a dataset of training shapes in order to model a shape probability distribution for use within a maximum a posteriori (MAP) estimation framework. Tsai et al.~\cite{tsai_shape-based_2003} took a similar approach by applying PCA to a training set of signed distance representation of shapes. More recent work has formulated the level set evolution as an energy functional minimization problem, where statistical information about shape and image features is incorporated by viewing the energy functional as the negative log of some probability density \cite{cremers_review_2007}. The probability density of image and shape features is modeled by employing, for example, Gaussian kernel density estimation \cite{cremers_efficient_2007}.

The work closest in spirit to our approach is not level set based, but is Van Ginneken's~\cite{van_ginneken_supervised_2006} machine learning extension of the standard region growing method. Traditionally, the region growing method recursively adds points to a growing region via a fixed rule (e.g., based on an image value threshold); however, Van Ginneken allows this rule to be learned from data, viewing the choice of whether or not to add a point as a binary classification problem that is solved via machine learning methods.

\subsection{Lung Nodule Image Segmentation}
\begin{figure}[!h]
    \centering
    \includegraphics[width=2in]{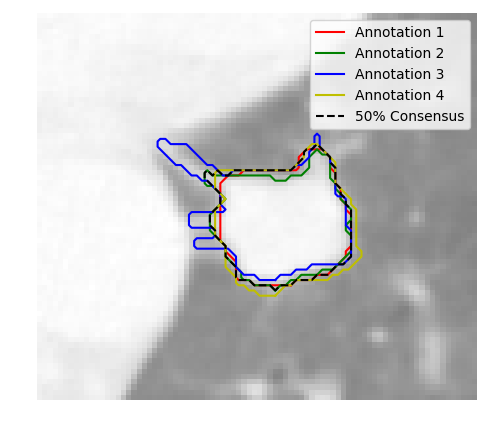}
    \caption{An annotated lung nodule from the LIDC dataset. Boundaries annotated by the four radiologists are shown in color, and the 50\% consensus consolidation of the four contours is shown in dashed black.}
    \label{fig:lidc-example}
\end{figure}

Many lung nodule segmentation works, including our own, leverage the publicly available LIDC dataset \cite{armato_iii_lung_2011} of lung CT data and radiologist annotations. The LIDC dataset contains 1018 lung CT scans that have been annotated by four radiologists (see Fig.~\ref{fig:lidc-example} for an example). Each radiologist visually examined each scan, and upon detecting a lung nodule, drew the perceived boundary of the lung nodule in each slice for which the detected nodule was present (according to that specific radiologist's informed opinion). These ``ground-truth" nodule boundary annotations, along with CT image volume data, are available in the LIDC dataset.

An assortment of methods have been applied to the lung nodule segmentation problem. Wang~et~al.~\cite{wang_central_2017} constructed a table of works, including their own, that reported the Jaccard overlap score. This score is also referred to as the ``intersection over union" (or IoU) score and is a measure of segmentation quality, defined as the size of the intersection between an algorithmic segmentation and a ground-truth segmentation, divided by the size of the union. The score ranges from zero to one, where zero indicates no overlap, and one indicates a perfect overlap.

Because we have also used the Jaccard overlap score as the measure of segmentation quality in our work, we have included and expanded this list of works in Table~\ref{tab:related_works_with_jaccard}, where we have placed our work in context for comparison with these other methods. Tachibana and Kido~\cite{tachibana_automatic_2006} combined a variety of image processing techniques such as thresholding, template-matching, and the watershed method (an edge-based method for determining boundaries), and obtained an average Jaccard overlap score of 0.5070 on 23 nodules. Wang~\cite{wang_segmentation_2009} used a dynamic programming approach and fusion method for combining information from multiple two-dimensional image slices. They reported an average Jaccard overlap score of 0.58 on 64 nodules. Messay~et~al.~\cite{messay_new_2010,messay_segmentation_2015} in their first work, applied a variety of morphological operations with a subsequent ``rule-based analysis", obtaining an average overlap of 0.63 over 68 nodules, whereas in their follow-up work they used a calibration process over training data to predict various thresholding and morphological parameters based on features computed from the image, improving their results to an average of 0.7170 over 66 nodules. Both Kubota~et~al.~\cite{kubota_segmentation_2011} and Lassen~et~al.~\cite{lassen_robust_2015} applied basic image processing techniques such as thresholding and morphological operations, as well as convexity information, achieving average Jaccard overlap scores of 0.69 and 0.52, respectively. Tan~et~al.~\cite{tan_segmentation_2013} used the watershed method, active contours, and Markov-random fields, achieving an average overlap score of 0.65 over a dataset containing 23 nodules. The work by Wang~et~al.~\cite{wang_central_2017} applied convolutional neural networks (CNNs) with a centrally-focused max-pooling operation to 493 test nodules, obtaining an average overlap score of 0.7116.

\begin{table}
    \centering
    \caption{Performance of various lung nodule segmentation methods under Jaccard and Dice similarity metrics when available.}
    \label{tab:related_works_with_jaccard}
    \begin{tabular}{llllcc}
    \toprule
    Authors                                               & Year & \multicolumn{2}{l}{Number of Nodules}  & Jaccard & Dice \\ \cline{3-4}
                                                          &      & Training & Testing & &  \\
    \midrule
    Tachibana and Kido~\cite{tachibana_automatic_2006}    & 2006 & -   & 23  & 0.5070~($\pm$0.2190) & 0.6729~($\pm$0.3593) \\
    Schildkraut et al.~\cite{schildkraut_level-set_2009}  & 2009 & -   & 23  & 0.4790 & 0.6477 \\
    Wang et al.~\cite{wang_segmentation_2009}             & 2009 & 23  & 64  & 0.5800 & 0.7342 \\
    Messay et al.~\cite{messay_new_2010}                  & 2010 & -   & 68  & 0.6300~($\pm$0.1600) & 0.7730~($\pm$0.2759) \\
    Kubota et al.~\cite{kubota_segmentation_2011}         & 2011 & -   & 23  & 0.6900~($\pm$0.1800) & 0.8166~($\pm$0.3051) \\
    Tan et al.~\cite{tan_segmentation_2013}               & 2013 & -   & 23  & 0.6500 & 0.7879 \\
    Farag et al.~\cite{farag_novel_2013}                  & 2013 & -   & 334 & N/A & N/A \\
    Lassen et al.~\cite{lassen_robust_2015}               & 2015 & -   & 19  & 0.5200~($\pm$0.0700) & 0.6842~($\pm$0.1308) \\
    Messay et al.~\cite{messay_segmentation_2015}         & 2015 & 300 & 66  & 0.7170~($\pm$0.1989) & 0.8352~($\pm$0.3318) \\
    Farhangi et al.~\cite{farhangi_3d_2017}               & 2017 & 488 & 54  & 0.5700~($\pm$0.1600) & 0.7261~($\pm$0.2759) \\
    Wang et al.~\cite{wang_central_2017} (level set)      & 2017 & 350 & 493 & 0.4350~($\pm$0.0952) & 0.6063~($\pm$0.1738) \\
    Wang et al.~\cite{wang_central_2017} (CNN)            & 2017 & 350 & 493 & 0.7116~($\pm$0.1222) & 0.8315~($\pm$0.2178) \\
    \midrule
    This work & & & & & \\
    \midrule
    Chan-Vese\cite{chan_active_2001} level set                        & -    & 672 & 112 & 0.2525~($\pm$0.2861) & 0.3229~($\pm$0.3530) \\
    U-net\cite{ronneberger2015u} CNN                        & -    & 672 & 112 & 0.6911~($\pm$0.1158) & 0.8104~($\pm$0.1044) \\
    Proposed LSML method                         & -    & 672 & 112 & {\fontseries{b}\selectfont 0.7185~($\pm$0.1114)} & {\fontseries{b}\selectfont 0.8362~($\pm$0.2178)} \\
    \bottomrule
    \end{tabular}
\end{table}

Relatively fewer works have applied the level set method for lung nodule image segmentation. Schildkraut~et~al.~\cite{schildkraut_level-set_2009} tested the level set method with energy terms, for ``increasing contrast of the segmented region relative to its surroundings", on 23 lung nodules in radiography images. They reported an average overlap using the Dice coefficient, $S(A,B) = \frac{2|A \cap B|}{|A| + |B|}$, of 0.6477 on the 23 lung nodules. The Dice score is related to the Jaccard score by $J = \frac{S}{2-S}$, and thus, the work by Schildkraut~et~al.~\cite{schildkraut_level-set_2009} reported an Jaccard overlap score of 0.4790. In the work of Tan~et~al.~\cite{tan_segmentation_2013}, which we have mentioned previously, a level set formulation of the geometric active contours method was employed as a post-processing step following an initial watershed segmentation, achieving an average Jaccard overlap score of 0.65. Farag~et~al.~\cite{farag_novel_2013} used a level set approach and incorporated an elliptical prior to aid in cases where lung nodules are in proximity to other anatomical objects. They reported a ``success rate" (where success is determined by visual inspection of the resulting segmentation) of 94.61\% on 334 lung nodules images from the LIDC dataset, but the Jaccard overlap score or similar measures of overlap are not reported. Farhangi~et~al.~\cite{farhangi_3d_2017} used the region-based Chan-Vese~\cite{chan_active_2001} active contour model to partition nodule and non-nodule regions based on region image-homogeneity using a level set formulation. In addition, a training set of nodule shapes was employed, and at each iteration during the level set evolution, the level set iterate was projected onto the linear span of the training shapes by solving a minimization problem that included a sparsity-inducing term to force coefficients in the weighted sum to be sparse. They used 542 lung nodules from the LIDC dataset, achieving an average Jaccard overlap score of 0.57 over a 10-fold cross-validation procedure. For comparison against their convolutional network model, Wang~et~al.~\cite{wang_central_2017} also applied a standard version (i.e., non-statistical and without specific tailoring to the lung nodule problem domain) of the region-based Chan-Vese level set model~\cite{chan_active_2001}, obtaining an average overlap score of 0.4350 over the same 493 nodules on which they tested their network model. In our work, we obtain an average Jaccard overlap score of 0.7185 over a testing dataset of 112 nodules using our machine learning extension of the level set method. We perform two additional experiments with our same data-split using the (non-machine learning) level set Chan-Vese\cite{chan_active_2001} method and the U-net CNN model\cite{ronneberger2015u}. Our methods and experimental configurations are described in more detail in Section~\ref{section:methods}.

\section{Methods}
\label{section:methods}
We provide motivation for the LSML method in Section~\ref{section:methods-motive}, and an algorithmic outline of the parameter tuning process in Section~\ref{section:methods-outline}. 
In Section~\ref{section:methods-dataprep}, we discuss how we prepared the LIDC dataset for use in this work. In Section~\ref{section:methods-init}, we describe our initialization routine, which yields a first guess of the segmentation given an image volume containing a lung nodule. In Section~\ref{section:methods-features} we describe the features that are used as inputs to the regression models in the LSML method used in our experiments.

\subsection{Motivation}
\label{section:methods-motive}
Consider a dataset of pairs $(M_l, c_l)$, where $M$ is an image and $c$ is a curve or surface annotating the boundary of the object to be segmented in $M$. As a preliminary definition, we note that the signed distance transform of a set $A \subset \mathbb{R}^n$ is defined as $d(x) = s(x) \inf_{y \in \partial A} \|x - y\|$ where $s(x) = 1$ if $x \in A$ and $-1$ otherwise. Now, in order to incorporate such labeled pairs $(M_l, c_l)$ to model the velocity term $\nu$ in Eq.~\eqref{eq:lspde}, we leverage an observation from Breen and Whitaker \cite{breen_level-set_2001}: if $\nu$ is the signed distance of the target boundary $c$, then the zero level set of $u$ converges to $c$ under the motion dictated by Eq.~\eqref{eq:lspde}. This occurs because the zero level set moves towards the boundary $c$ with speed equal to the distance from the boundary, where the choice of expansion or contraction of the zero level set is controlled by the sign in the signed distance.

Although setting $\nu$ to the signed distance transform of the target boundary $c$ assumes the solution, these observations strongly suggest an approximation scheme of the form $\nu \approx V$, where $V$ is obtained from a machine learning regression model. Using image and shape data as inputs, the regression model $V$ is calibrated to approximate the signed distance values from the target boundaries provided by the labeled dataset. Thus, the level set evolution is guided towards more statistically likely segmentation configurations in cases where the correct segmentation solution is unknown. This is the essence of the LSML approach.

\subsection{Outline of the LSML method}
\label{section:methods-outline}
First, Eq.~\eqref{eq:lspde} is discretized using the upwind scheme of Osher and Sethian~\cite{osher_fronts_1988}. Next, the velocity term $\nu$ in the discretization is instead replaced with a sequence of approximations by a regression models, i.e., $\nu \approx V^n$ for $n=0, 1, \ldots$, thus resulting in the iteration,
\begin{equation}
    \label{eq:lsapprox}
    u^{n+1}_{ijk} = u^n_{ijk} + \Delta t \, V^n_{ijk} \, \nabla^n_{ijk}
\end{equation}
The input to the regression model $V^n$ is a feature vector $F$ computed from image and shape data, and thus $V^n_{ijk} = V^n_{ijk}\left( F_{ijk} \right)$, where $F = F(u^n, M)$. In other words, the feature vector, which is the input to the regression model $V^n$, is a function of the image and corresponding level set iterate, from which image and shape features can be extracted. As a simple example, consider a two-component feature map function that yields two features, $F_{ijk}(u, M) = \left[ \sum_{qrs} H(u_{qrs}), \; M_{ijk} \right]$. The first feature, which approximates the volume enclosed by the zero level set of $u$, does not depend on any particular local $(i,j,k)$ position (since we are summing over all positions in the image volume) and is thus a global shape feature. The second feature $M_{ijk}$ depends on both the local spatial grid index $(i,j,k)$ and the image $M$, and is therefore a local image feature. Generally, the feature map function can consist of an assortment of combinations of local, global, image, and shape features.

\begin{figure}
	\centering
	\includegraphics[width=0.4\textwidth]{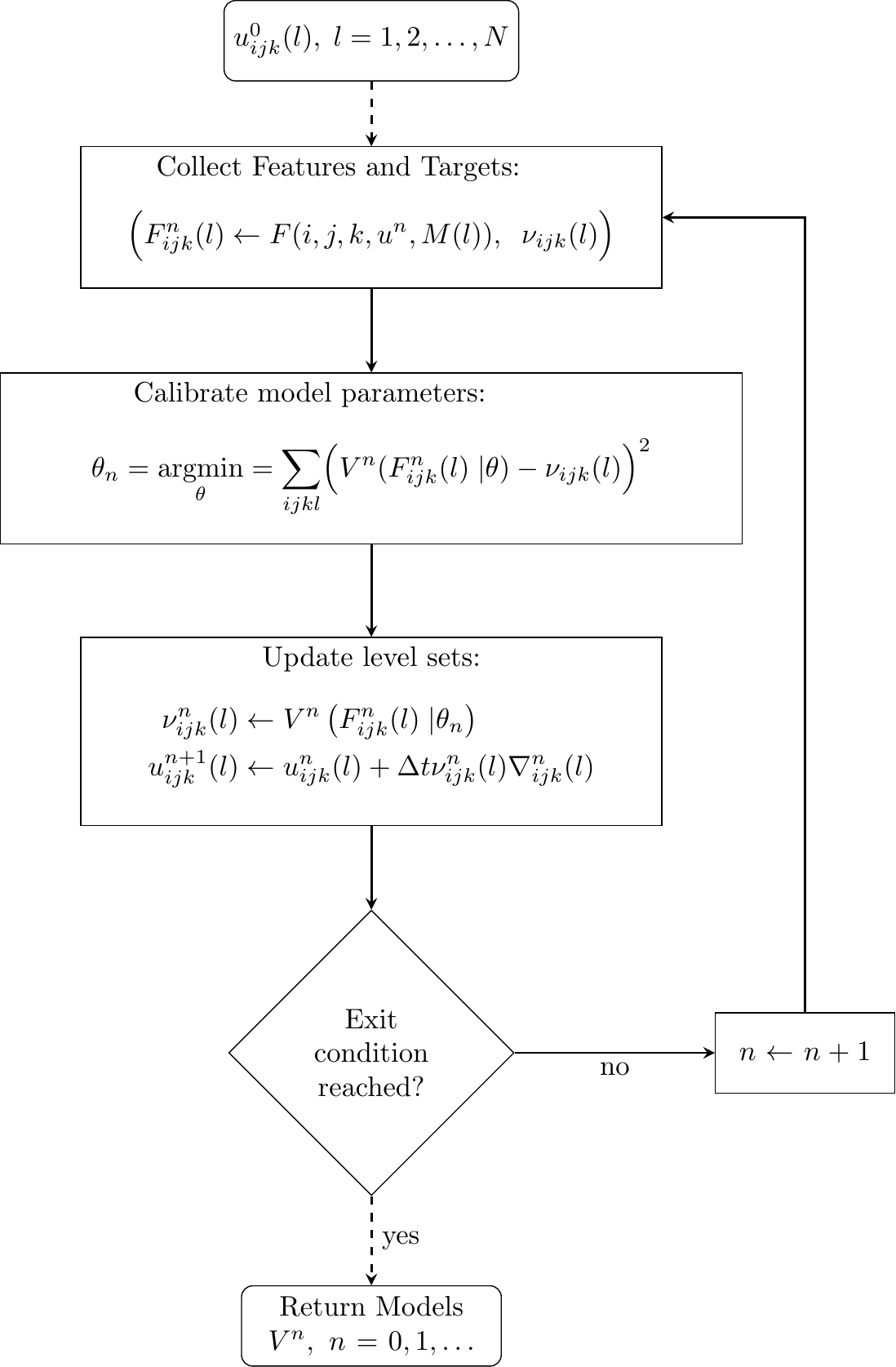}
	\caption{Flow diagram of the training process of the LSML method}
	\label{fig:lsml_training_flow}
\end{figure}

The goal of the regression model $V^n$ is to learn a mapping from the feature vector $F_{ijk}$ (whose components provide local, global, image, and shape information) to the signed distance value at $(i,j,k)$ because, as discussed in Section~\ref{section:methods-motive}, these values guide the evolving zero level set towards the target shape. In other words, we assume that the direction and speed of the evolving segmentation boundary can be written as a function of local, global, image, and shape information. The regression models $V^n$ are obtained sequentially in the training process outlined below and in the flow chart in Fig.~\ref{fig:lsml_training_flow}, assuming a dataset $(M(l), c(l))_{l=1}^N$ of $N$ images and their ground-truth boundaries:

\begin{enumerate}
    \item Initialize each $u^0(l)$ according to a computationally inexpensive process that produces a first approximation of each segmentation for image $M(l)$. Set $n = 0$.
    \item Calibrate the parameters of the $n$th regression model by least squares over all spatial coordinates and examples. Specifically, determine the optimal parameters $\theta^n$ for model $n$ via:
    $$
    	\theta^n = \argmin_\theta \sum_{ijkl} \left(V^n_{ijk}(\cdot | \theta) - \nu_{ijk}(l)\right)^2
    $$
    Note that we have suppressed the argument to $V^n$. In full, $V^n$ is a function of $F(u^n(l), M(l))$, i.e., of the feature vector for the current level set iterate $u^n$ and of the image $M$ for the $l$th example.
    \item Step each $u^n(l)$ forward in time by the level set evolution PDE discretization given by Eq.~\eqref{eq:lsapprox}
    \item If segmentation quality metrics (e.g., the intersection over union score) that are observed over a separate validation set have not reached a termination condition (e.g., decreasing over the most recent steps), then set $n \leftarrow n+1$, and go to Step 2, unless some maximum allowable iteration has been reached.
\end{enumerate}

The training procedure yields a sequence of regression models $V^n$ that can be deployed on new, unseen images using the iteration in Eq.~\eqref{eq:lsapprox}. For our models $V^n$ we employ the Random Forest regression~\cite{breiman_random_2001} algorithm (as implemented in the Scikit-Learn library for Python~\cite{sklearn}) with one hundred decision tree regressors in the forest. We also experimented with neural networks~\cite{friedman_elements_2001} and support vector regression~\cite{friedman_elements_2001} but found that the Random Forest model out-performed these approaches in preliminary experimentation in terms of their accuracy and training time. Lastly, we note that both the training procedure above, as well as the method when deployed on unseen testing data, can be made more computationally efficient by considering spatial coordinates only in a narrow band about the evolving zero level set for the segmentation. Fast computational methods exist to determine this narrow band region~\cite{adalsteinsson_fast_1995}.

\subsection{Data preparation}
\label{section:methods-dataprep}
We use the data provided in the LIDC dataset\cite{armato_iii_lung_2011} for our experiments. The data undergoes a number of pre-processing steps, which we first describe briefly and then in more detail in the subsequent subsections. First, the data is analyzed to obtain only nodules where all four radiologist annotators agree upon the existence of a lung nodule at a particular location in the scan. This yields 896 nodules. Next, these annotation contours are converted to boolean-valued target volumes, and multiple annotations for the same nodule are consolidated into a single boolean volume. Afterward, for each of the 896 lung nodules selected, we standardize the image volumes containing the lung nodule, as well as its associated ground-truth boolean-valued volumes, to have uniform voxel spacing of one millimeter, because the CT scans in the LIDC data have been generated with different scanning devices and scanner parameters, resulting in different image volume resolutions. Lastly, we randomly partition this dataset of 896 lung nodules and the respective ground-truth segmentations into subsets of size 672, 112, and 112 for training, validation, and testing, respectively.

\subsubsection{Gathering nodules with four annotators}
The CT image volumes from the LIDC dataset were annotated by up to four radiologists, but the physical lung nodules lack a universal identifier. By ``annotation", we mean the sequence of planar curves in each CT slice describing the boundary of a particular lung nodule, as determined by a single radiologist. Symbolically, a lung nodule annotation can be written $\mathcal{A} = ( C_j, C_{j+1}, \ldots, C_{j+n} )$, where $C_k$ is the curve describing the nodule boundary in image-slice $k$ of the CT volume. In our experiments, we only use the lung nodules that have been detected and annotated by all four radiologists. 

Thus, we begin by estimating when multiple annotations refer to the same physical nodule in an image. We accomplish this by first defining a distance function, $d$, that describes the nearness of two nodule annotations, $\mathcal{A}_i$ and $\mathcal{A}_j$. Next, for a given scan, we compute a distance matrix, $D$, where the $(i,j)$ entry is $d(\mathcal{A}_i, \mathcal{A}_j)$ (i.e., the distance from annotation $i$ to annotation $j$), thus providing the pair-wise distances between all annotations in the scan. Two annotations are said to be adjacent when $D_{ij} \leq \tau$, where $\tau$ is a threshold parameter that is empirically determined. The value of $\tau$ is first initialized to be equal to the slice thickness (in millimeters) of the particular scan, which is a parameter that can be found in the DICOM image data in the LIDC dataset. Nodule annotations are said to refer to the same physical nodule in the scan when they belong to the same connected component of the adjacency graph, which is formed by thresholding the pair-wise distance matrix $D$ by $\tau$. If afterwards there are annotation groupings with size greater than four nodules (i.e., greater than the number of annotating radiologists), we reduce the threshold parameter $\tau$ by a multiplicative factor, and we repeat the process. In our work, we find that the distance function, $d$, between two annotations that takes the minimum over all pairwise 3D distances between the coordinates of the two annotations works well, which is confirmed visually. From this process, we obtain 896 lung nodules, each having annotations by exactly four radiologists. This approach for clustering annotations is implemented in the \texttt{pylidc}~\cite{hancock_lung_2016} Python package.

\subsubsection{Volume interpolation}

The pixel and slice spacing (i.e., the within- and between-slice scan resolutions, respectively) varies among scans in the LIDC dataset, and to normalize this we construct bounding boxes about each nodule, which we then interpolate to have uniform voxel spacing across all scans. For each lung nodule, of the 896 obtained with four annotations (discussed in the previous section), a common reference frame for the four associated annotations is formed. In this common reference frame, we convert each annotation into a boolean-valued volume that is one inside each annotating contour and zero outside, using the ray-casting method implemented in the \texttt{matplotlib}~\cite{Hunter:2007} Python package. Next, we perform a trilinear interpolation on both the image and the four associated boolean-valued volumes so that the resulting volumes are isotropic with a one-millimeter spacing between voxels. The volumes are interpolated so that each volume is 70$^3$ cubic millimeters, which was chosen to account for the nodule with the largest observed diameter of 60 millimeters and to leave sufficient padding of non-nodule voxels about every nodule. Thus, the interpolated volumes are of dimensions, $71 \times 71 \times 71$.

\subsubsection{Consolidation of multiple ground truths and final pre-processing steps}

For each nodule, we consolidate its four ground-truth segmentations (i.e., the boolean-valued indicator volumes of the lung nodule), $B^k$, $k=1,2,3,4$, into a single ground-truth segmentation $B^*$ by computing
$
B^* = \argmax_{B\in\Omega} \left\{ \frac{1}{4}\sum_{k=1}^4 J\left(B, B^k\right) \right\}, \;\; \Omega = \{0,1\}^{71 \times 71 \times 71}
$
where $J(\cdot, \cdot)$ is the Jaccard overlap function. The quantity $B^*$ is sometimes referred to as the ``Jaccard median"\cite{spath_minisum_1981,watson_algorithm_1983,chierichetti_finding_2010} and is the best consolidation of the four annotations in the sense that it agrees most with all four annotations under the Jaccard overlap measure. Although we use the Jaccard median for our experiments, we observe that the typically-used 50\% consensus consolidation (where the consolidated segmentation is set equal to one where 50\% or greater agreement occurs in the nodule's segmentations and is set to zero otherwise) is often in high agreement with the Jaccard median. For example, in our case, $J(B_{50}, B^*) \approx 0.96$ on average, where $B_{50}$ denotes the 50\% consolidation. This, combined with its simplicity (whereas the Jaccard median requires numerically solving a constrained optimization problem), justifies the use of the 50\% consensus consolidation method that is often used in other works that use annotations from the LIDC dataset.

As the final pre-processing step, we standardize each image volume by subtracting off each respective mean and dividing by each respective standard deviation. This process results in datasets, $\Bigl(M(k), B(k)\Bigr)_{k=1}^N$, of image and segmentation pairs, where $N=672$ for the training dataset and $N=112$ for the validation and testing datasets.

\subsection{Initialization}
\label{section:methods-init}
\begin{figure}[hp]
	\centering
	\begin{subfigure}[b]{\textwidth}
		\centering
		\includegraphics[width=0.9\textwidth]{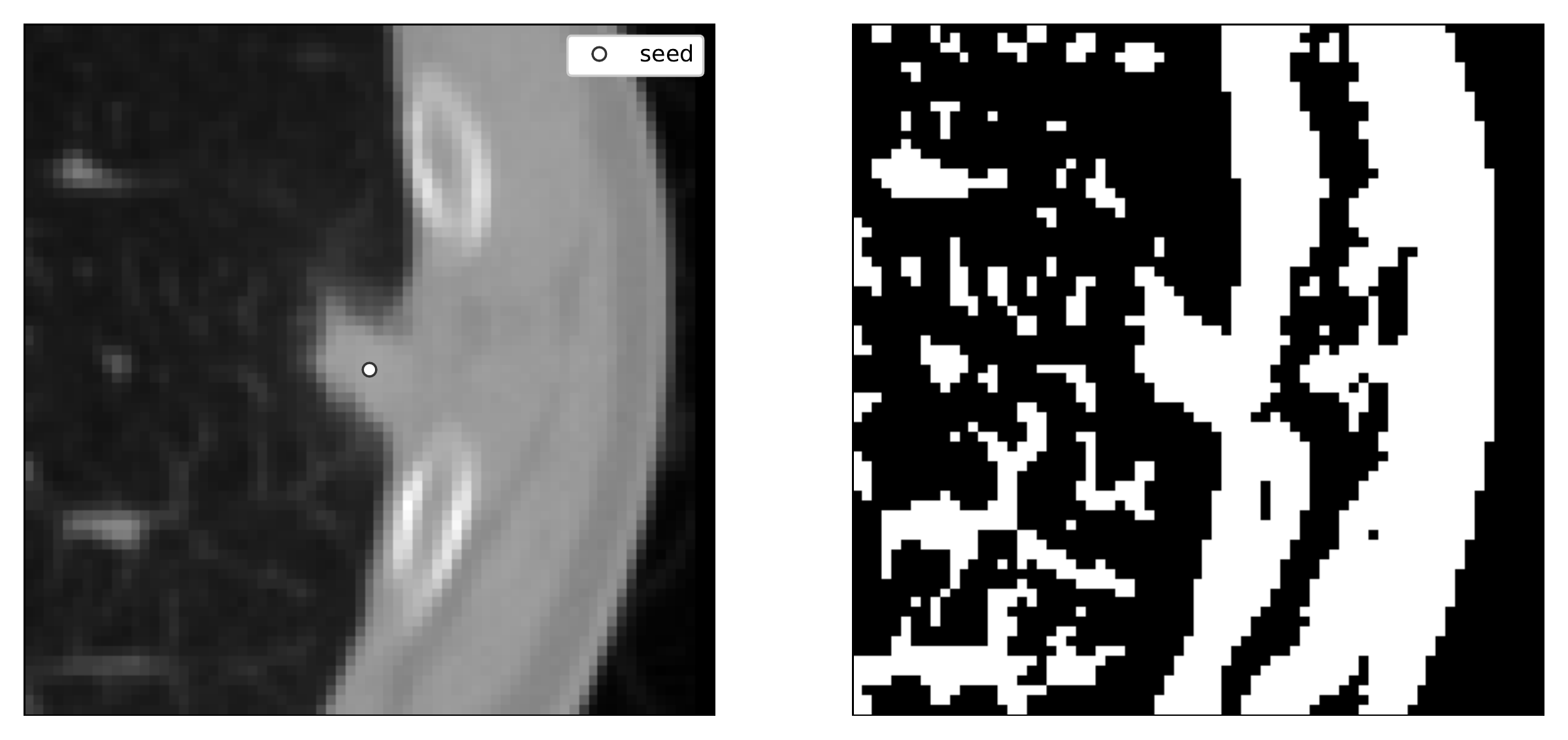}
		\caption{{\bf Left:} The center (35th) slice of the image volume is shown, as well as the center point of the image, indicated by the ``seed" label. {\bf Right:} The image is thresholded by comparing the value at each voxel to the weighted average of the neighboring values, and all connected components not touching the seed point are eliminated.}
		\label{fig:lidc_level_set_init1}
	\end{subfigure}
	
	\begin{subfigure}[b]{\textwidth}
		\centering
		\includegraphics[width=0.9\textwidth]{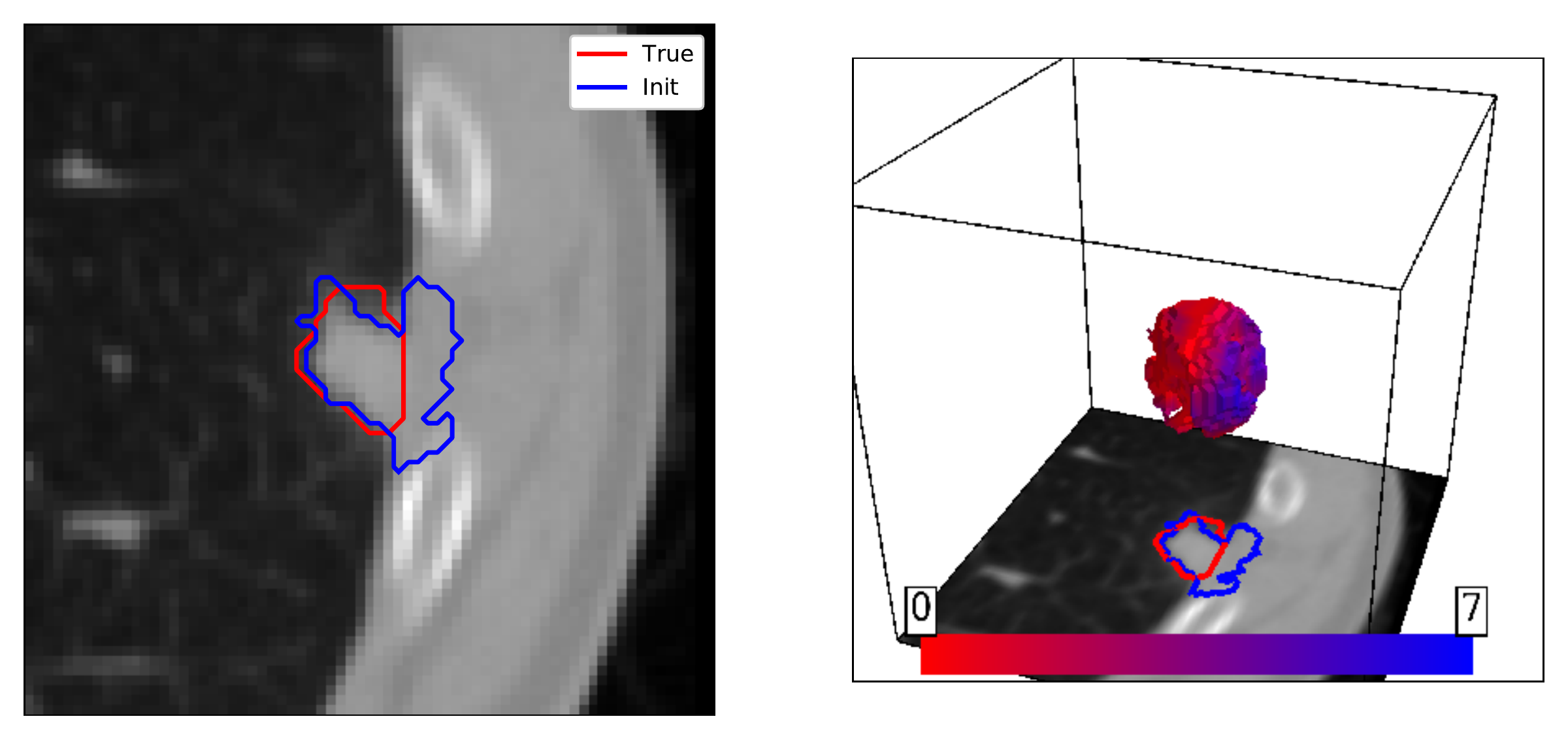}
		\caption{{\bf Left:} Radii are sampled by sampling azimuth and zenith angles, and computing the distance until the background value (i.e., a value of zero in the binary image volume) is first reached from a ray beginning at the seed point. Radii beyond the 70th percentile of the observed radii are trimmed to obtain the initialization shown in the blue curve in the center slice. The red curve shows the ground truth in this slice, for comparison. {\bf Right:} The initialization volume is shown in red and blue hues, where the color on the surface indicate the distance from the ground truth. Red indicates a distance of zero, and blue indicates the maximal distance of 7 voxels. The color-to-distance encoding is shown in the color bar in the figure. The center (35th) slice of the image volume is shown below the surface.}
		\label{fig:lidc_level_set_init2}
	\end{subfigure}
	
	\caption{Example of the level set initialization procedure for lung nodule images.}
	\label{fig:lidc_level_set_init}
\end{figure}
In devising a segmentation initialization procedure, we use the observation that lung nodules are often, but not always, dense (thus appearing discernibly brighter than their surroundings in CT scan image slices), and are approximately spherical in shape. These assumptions about the shape and appearance of the nodules are frequently violated, as they may not appear brighter than their surroundings and may have irregularly shaped margins far from spherical. Nevertheless, the goal in initialization is only to provide a reasonable enough guess, so as to allow the LSML algorithm to improve upon it.

The initialization process uses local thresholding, connected component analysis\cite{scipy}, and a ray-casting based post-processing method to trim extraneous radii. An example is shown in Fig.~\ref{fig:lidc_level_set_init}. Local thresholding of the image yields regions in the image that have similar image intensity values. Connected component analysis removes extraneous binary components obtained by thresholding, except the component that is closest to the seed point. Finally, the radius trimming technique serves to remove any parts of the binary component that extend beyond a specified distance from the center of the volume. This initialization process involves two free parameters, a smoothing factor $\sigma$ and a radius percentile value $p_r$, that are calibrated by performing a grid search over the training data.

In more detail, the procedure begins by convolving a Gaussian smoothing kernel $G_\sigma$ (with parameter $\sigma$) with the image, and the image is then thresholded by setting values that are larger than the smoothed value to one, and those that are below, to zero. This produces a boolean valued volume. Afterwards, we determine the connected component of this binary image volume that is closest to the center point of the volume. All pixel values in the binary image that are not in this connected component are set to zero. Next, in spherical coordinates with the center of the image as the origin, we sample azimuth and zenith angles uniformly and determine the corresponding radius, defined as the distance until the first voxel where a ray emanating from the center of the volume meets a value of zero in the boolean volume. After sampling many azimuth and zenith angles, we compute $p_r$, the percentile of the radii observed. Any ray with a radius extending beyond $p_r$ is trimmed (see Figure~\ref{fig:lidc_level_set_init2}, left, in blue), i.e., the boolean image volume is set to zero outside a sphere of radius $p_r$. If after trimming the radii, there are multiple connected components, we then choose the component closest to the seed point, and set the others to zero. The initial values of $u^0$ are set to $+1$ where the final boolean initialization is $1$, and to $-1$ where the boolean initialization is $0$. Thus, the initial segmentation boundary given by the zero level set of $u^0$ is set to match the boundary of the boolean valued volume from this initialization procedure.

The free parameters, $\sigma$ and $p_r$, are determined through a grid search procedure over the training data. Specifically, we search over the parameter values, $\sigma \in \{1, 2, \ldots 7\}$ and $p_r \in \{50, 55, \ldots, 80\}$. For each parameter combination in the Cartesian product of these sets of values, we compare this segmentation from our initialization procedure against the ground truth segmentation with the Jaccard overlap score. The parameter combination with the highest average overlap score over the training data is used, which we determine to be $\sigma=4$ and $p_r=70$.

\subsection{Features used}
\label{section:methods-features}
The definitions of all the features used as components of the feature vector $F$ used in our experiments are given in Appendix~\ref{section:lsml_features}. We employ two feature sets that we call Feature Map~1 and Feature Map~2, which we describe here.

The first set of features that we apply to the lung nodule image segmentation problem, which we call Feature Map~1, is described in Table~\ref{tab:fmap1}. These features are simple and generic and serve as a baseline against which to compare Feature Map~2, an extension of Feature Map~1. Feature Map~1 contains 10 shape descriptors, where segmentation moment features are computed at two orders and along the three coordinate axes. Image features (of which there are 4) are computed at two Gaussian-smoothing scales, a fine scale ($\sigma=0$) and a coarse scale ($\sigma=3$), for a total number of 8 image features. Thus, Feature Map~1 comprises a total of 18 features.

\begin{table}[h]
	\centering
	\caption{Feature Map 1: Image features are computed at fine ($\sigma=0$) and coarse ($\sigma=3$) scales. The symbols $\Omega = \{x : u > 0\}$ and $\partial\Omega = \{x : u = 0\}$ are used.}
	\label{tab:fmap1}
	\begin{tabular}{llcc}
		\toprule
        \# & Feature & Local & Global \\
		\midrule
        1 & Volume $= |\Omega|$ & & \checkmark \\
		1 & Surface area $= |\partial\Omega|$ & & \checkmark \\
		1 & Isoperimetric ratio $= 36\pi\frac{|\Omega|^2}{|\partial\Omega|^3}$ & & \checkmark \\
        6 (3 axes$\times$2 orders) & Moments of $\Omega$, order = 1,2 & & \checkmark \\
		1 & Distance from $(i,j,k)$ to center of mass & \checkmark & \checkmark \\
		2 (1$\times$2 $\sigma$'s) & Image average in $\Omega$ & & \checkmark \\
		2 (1$\times$2 $\sigma$'s) & Image variability in $\Omega$  & \checkmark \\
		2 (1$\times$2 $\sigma$'s) & Image value at $(i,j,k)$ & \checkmark & \\
		2 (1$\times$2 $\sigma$'s) & Image edge at $(i, j, k)$ & \checkmark & \\
		\bottomrule
	\end{tabular}
\end{table}

The features in Feature Map~2 extend Feature Map~1 and are listed in Table~\ref{tab:fmap2}. Many of the features involve both local and global aspects for their computation. These additional global-local (or ``glocal") features are local in the sense that they require the local voxel coordinate $(i,j,k)$, but use previously computed global features such the ``center of mass" in their computation. Feature Map~2 includes a total of 109 features, and numbers of the contributed features are provided in Table~\ref{tab:fmap2}. The ``distance to center of mass" statistical features in Feature Map~2 (i.e., the average, variability, and maximum) supplement the ``distance to center of mass" feature from Feature Map~1 with a more global context. The ``slice areas" and ``slice areas absolute change" features are computed by calculating the areas of each slice through $\Omega$ (the interior of the nodule's segmentation boundary) in the three axes directions. These provide a more localized version of the purely global volume feature in Feature Map~1.

\begin{table}[h]
	\centering
	\caption{Feature Map 2: Image features are computed at fine ($\sigma=0$) and coarse ($\sigma=3$) scales. The symbols $\Omega = \{x : u > 0\}$ and $\partial\Omega = \{x : u = 0\}$ are used.}
	\label{tab:fmap2}
	\begin{tabular}{llcc}
		\toprule
		\# & Feature & Local & Global \\
		\midrule
		18 & All Feature Map~1 features (see Tab.~\ref{tab:fmap1}) & \checkmark & \checkmark \\
		2  (1$\times$2 $\sigma$'s) & Image average over $\partial\Omega$ 				  &			   & \checkmark \\
		1  & Distance to center of mass average 					& 				  & \checkmark \\
		1  & Distance to center of mass variability &  & \checkmark \\
		1  & Distance to center of mass maximum &  & \checkmark \\
		3  (1$\times$3 axes) & Slice areas & \checkmark & \checkmark \\
		3  (1$\times$3 axes) & Slice areas absolute change & \checkmark & \checkmark \\
		40  (20$\times$2 $\sigma$'s) & Image samples along normal & \checkmark & \checkmark \\
		40  (20$\times$2 $\sigma$'s) & Image samples along ray to center of mass & \checkmark & \checkmark \\
		\bottomrule
	\end{tabular}
\end{table}

\begin{figure}
	\centering
	\includegraphics[width=2in]{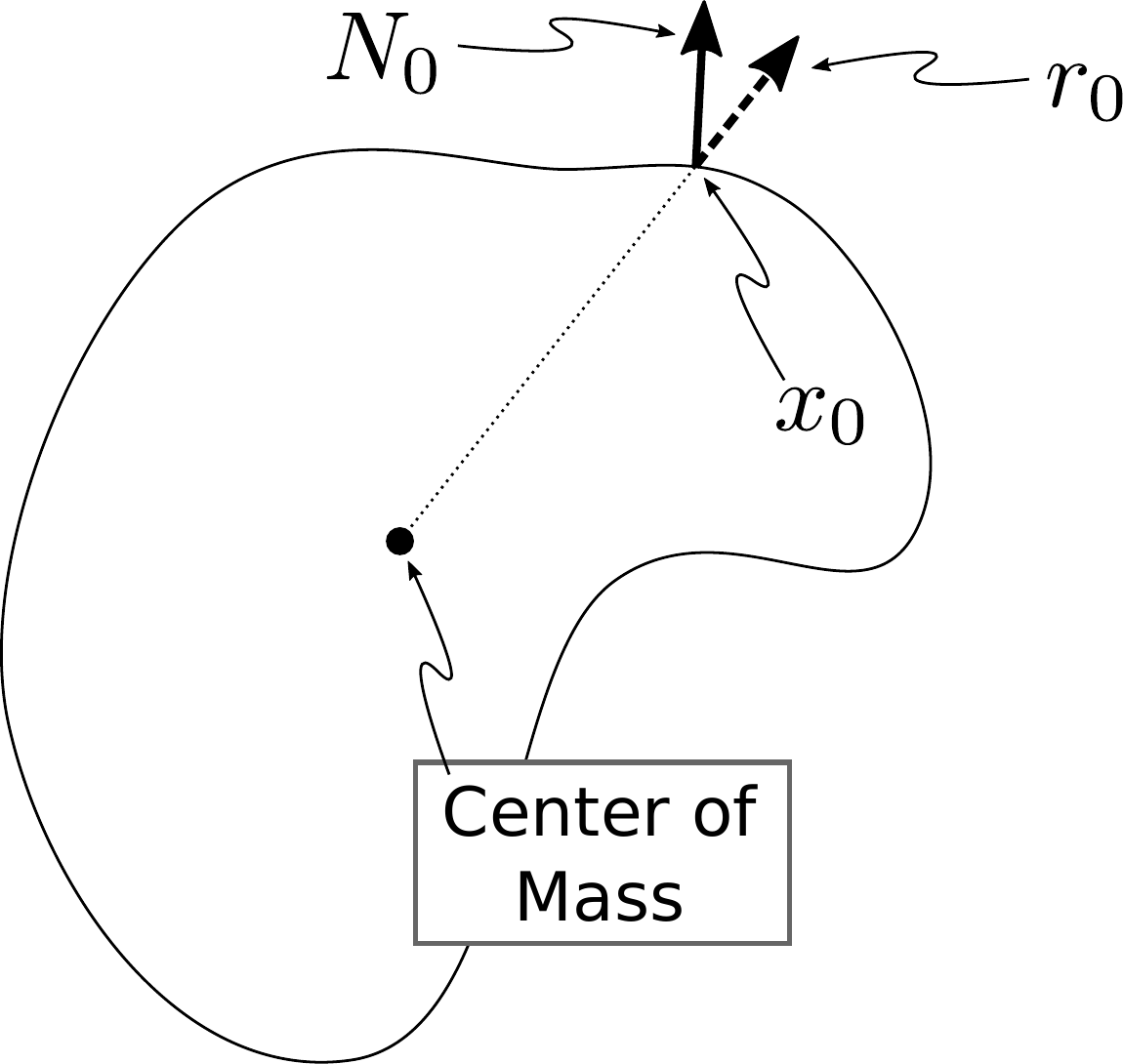}
	\caption{Schematic of the features from Feature Map~2 that sample the image along the normal $N_0$ and direction of the line segment connecting the center of mass of $\{x: u > 0\}$ and the position $x_0$.}
	\label{fig:sample-features}
\end{figure}

The ``image samples along normal" and ``image samples along ray to center of mass" features, illustrated in Fig.~\ref{fig:sample-features}, provide context of the image along two lines emanating from a given coordinate $x_0 = (i,j,k)$. Image values are sampled along two lines: (1) in the direction of the unit normal at $x_0$, and (2) in the direction of the line segment connecting the center of mass and $x_0$. For each of these lines, 10 samples are taken in both the inward and outward directions.

\subsection{Additional experiments}
In addition to our comparisons against other work from the literature (see Table~\ref{tab:related_works_with_jaccard}), we also perform two additional experiments using established methods over our same data-split and preparation process described in \ref{section:methods-dataprep}. Specifically, we implement the Chan-Vese\cite{chan_active_2001} segmentation model, a non-machine-learning based level set method and the U-net CNN\cite{ronneberger2015u} model. In this section, we describe the setup and configurations for these experiments.

Although the Chan-Vese segmentation method is not optimizable in the machine learning sense, it does involve a number of hyper-parameters that we optimize via a grid-search over the training data. In particular, we perform a grid search over the smoothing parameter $\sigma$ and parameters, $\lambda_1$ and $\lambda_2$, which weight the outer and inner regions of the loss functional. The initial level set is also a free choice in the Chan-Vese method, for which we employ the initialization procedure described in \ref{section:methods-init} so as to provide a more impartial comparison against our LSML method.

We follow the U-net architecture but with minor modifications related to the input and output dimensions. The original U-net architecture was designed for 512x512 gray-scale or color-channeled images. Recall that our image volumes are cubic of size 71 pixels along each axis. Firstly, we treat the z-axis as the ``color channel" axis with respect to the network input, which means that the two-dimensional convolutions occur with respect to the first two axes of the input image volume, i.e., with respect to the transverse anatomical plane. Secondly, we apply reflection padding to the input volumes along the first two axes such that the resulting image volume is of dimension 80x80x71. This is done because the U-net architecture involves four applications of down-sampling (via max-pooling) and four applications of upsampling, both by a factor of two. Padding the input image from 71 to 80 ensures that the output volume is the same dimensions of the input volume. Besides these dimensional modifications, we follow the same architecture and design as the original U-net\cite{ronneberger2015u} in terms of number of feature maps, convolutional kernel sizes, and activation functions. The network is implemented using TensorFlow\cite{tensorflow2015-whitepaper} and is trained for 500 epochs using the ADAM optimization method\cite{kingma2014adam}.

\section{Results}
\label{section:results}
\begin{figure}[htb]
	\centering
	\begin{subfigure}[t]{0.4\textwidth}
		\centering
		\includegraphics[width=\textwidth]{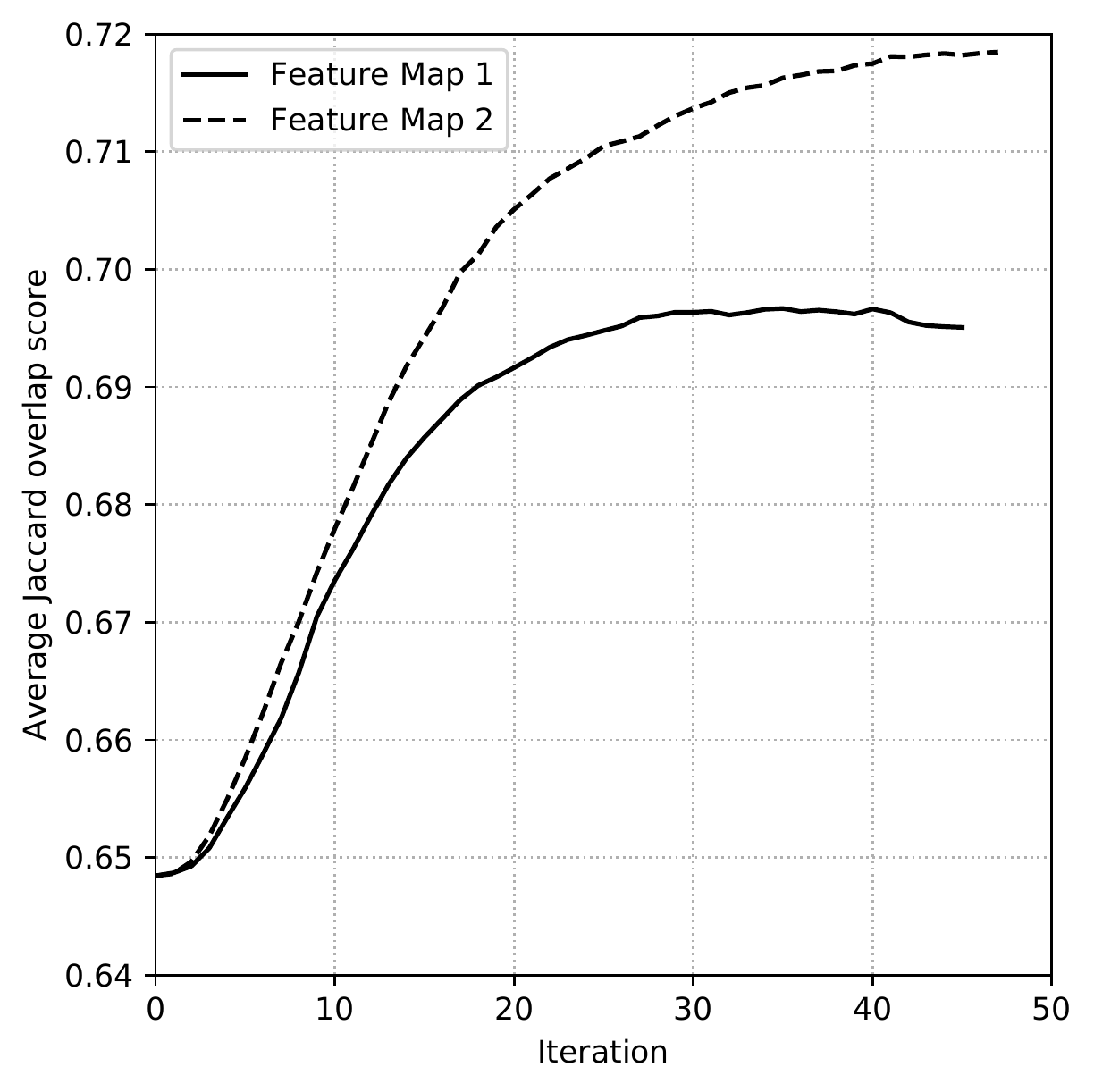}
		\caption{Average Jaccard overlap scores over the testing dataset against iteration number for Feature Maps~1~and~2.}
		\label{fig:lidc_overlap_iter}
	\end{subfigure}\hfill%
	\begin{subfigure}[t]{0.4\textwidth}
		\centering
		\includegraphics[width=\textwidth, height=3in, keepaspectratio]{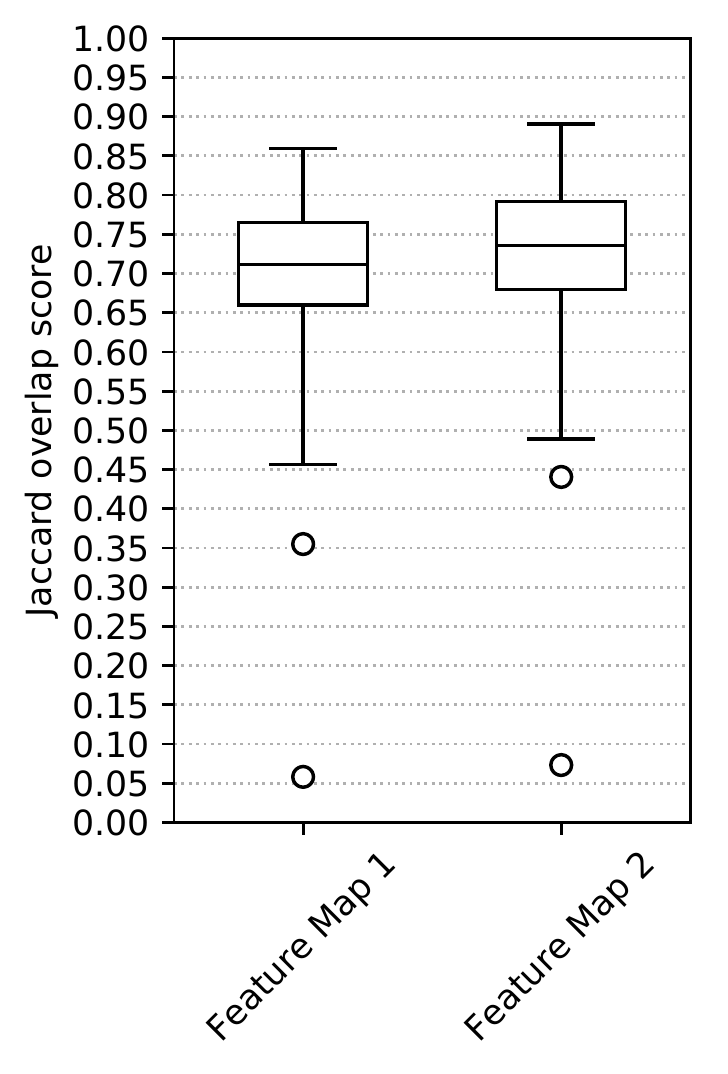}
		\caption{Box plots of the distribution in Jaccard overlap score for Feature Map~1 and Feature Map~2.}
		\label{fig:lidc_boxplots}
	\end{subfigure}
	\caption{Results for the LSML method applied to lung nodule image volumes from CT scans in the LIDC dataset.}
	\label{fig:lidc_results}
\end{figure}

\begin{table}[h]
	\centering
	\caption{Results for Feature Map~1 and Feature Map~2. $n^*$ is the iteration number at which the LSML algorithm terminates for the validation dataset, and $\bar{J}$ is the average Jaccard overlap score over the testing dataset that was achieved at the respective iteration $n^*$.}
	\label{tab:overlap}
	\begin{tabular}{ccc}
		\toprule
		& $n^*$ & $\bar{J}$ \\
		\midrule
		Feature Map~1 & 45 & 0.6951~($\pm$0.1119) \\
		Feature Map~2 & 47 & {\fontseries{b}\selectfont 0.7185~($\pm$0.1114)} \\
		\bottomrule
	\end{tabular}
\end{table}
In Table~\ref{tab:overlap} we report the average Jaccard overlap scores, obtained for the testing dataset, along with the optimal iteration from the validation dataset. In Fig.~\ref{fig:lidc_overlap_iter}, we plot the average Jaccard overlap score over the testing dataset against the iteration number for Feature Map~1 (shown in solid black) and Feature Map~2 (shown in dashed black). At initialization, the average Jaccard overlap score is 0.6484~($\pm$0.1119), which is higher, or comparable, to the scores obtained by some of the methods listed in Table~\ref{tab:related_works_with_jaccard}. Using Feature Map~1, the LSML algorithm terminates after 45 iterations, obtaining a final average Jaccard overlap score of 0.6951~($\pm$0.1127), an increase of 7.2\% relative to the average overlap score at initialization. With Feature Map~2, the algorithm terminates after 47 iterations, and obtains a final average overlap score of 0.7185~($\pm$0.1114), a 10.8\% relative increase over the average score at initialization. Thus, Feature Map~2 produces a 3.4\% relative increase in overlap score over the overlap score produced under Feature Map~1, which is a modest, but significant ($p=2\times10^{-6}$) increase, indicating the usefulness of the ``glocal" features that are included in Feature Map~2. We also observe from Figure~\ref{fig:lidc_overlap_iter} that the score increases more rapidly during the earlier iterations than the latter ones, indicating that the more substantial changes in the segmentation shape (in the sense of those that produce relatively larger increases in the overlap score) occur in the early iterations, whereas the latter iterations serve in making relatively smaller refinements to $\bar{J}$, as the slopes of the two curves show.

In Figure~\ref{fig:lidc_boxplots}, the distributions of overlap scores for the Feature Maps~1~and~2 are given on the left and right, respectively. The median overlap score for Feature Map~1 is 0.7115 (which is slightly greater than the mean overlap of 0.6951), and the 25th percentile to the 75th percentile stretches from 0.6596 to 0.7650, thus giving an inter-quartile range of 0.1054. For Feature Map~2, the median overlap score is 0.7356 (also slightly larger than the mean overlap of 0.7185), with the 25th and the 75th percentiles at 0.6793 and 0.7918, respectively, having a slightly larger inter-quartile range than Feature Map~1 of 0.1125. The large range in overlap score is due to the varying difficulty in segmenting different types nodules, with some being significantly more difficult to segment than others (which is discussed in Section~\ref{subsection:results_by_type}).

\begin{figure}[htb]
	\centering
	\includegraphics[width=0.9\textwidth]{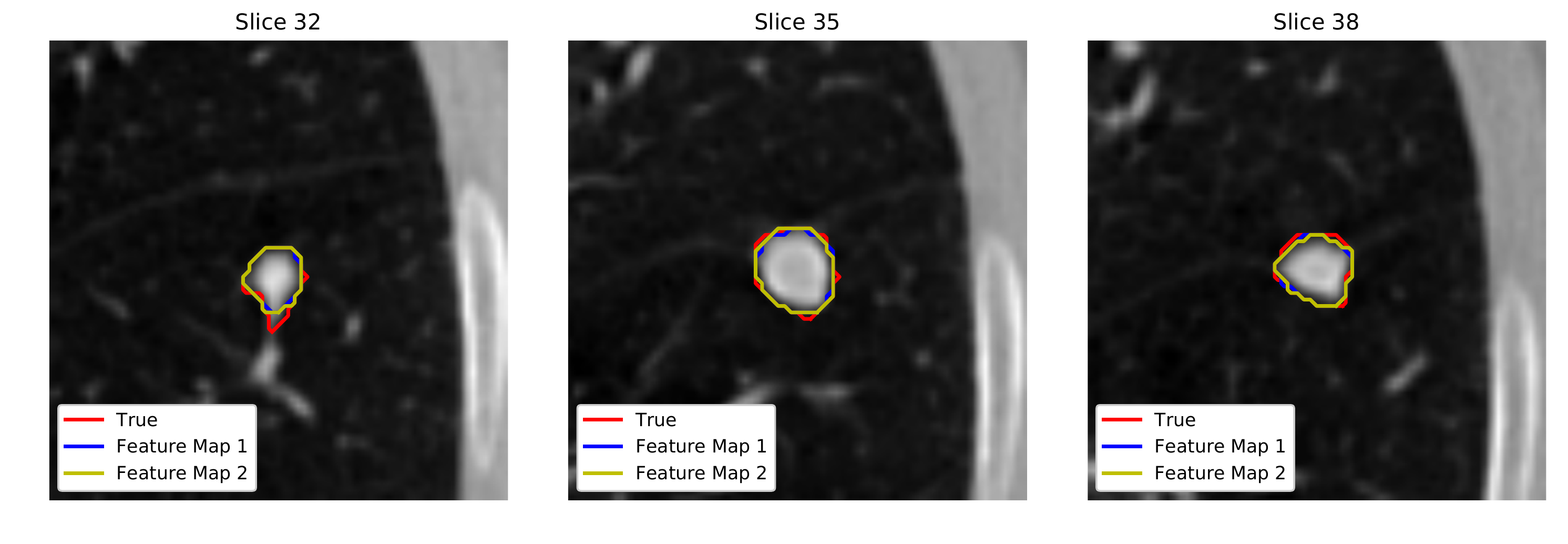}
	\caption{The lung nodule with maximal overlap between the true boundary (red) and its segmentation occurring under Feature Map~2 (yellow), which is second highest overlap occurring under Feature Map~1.}
	\label{fig:good_nodule}
\end{figure}

The maximal overlap scores observed for Feature Maps~1~and~2 are 0.8593 and 0.8906, respectively. The lung nodule with the highest overlap score under Feature Map~2, and the second highest overlap score (of 0.8573) under Feature Map~1, is shown in Figure~\ref{fig:good_nodule}. It is an isolated nodule with a well-defined boundary, as one might expect. Both feature maps yield segmentations that capture the true boundary of this nodule relatively well, with the most notable difference occurring in the 32nd slice, where the true boundary includes slightly more of the region of the nodule that is in close proximity to the vasculature in the posterior direction (i.e., near the bottom of the nodule).

\begin{figure}
	\centering
	\begin{subfigure}[t]{0.45\textwidth}
		\centering
		\includegraphics[width=\textwidth]{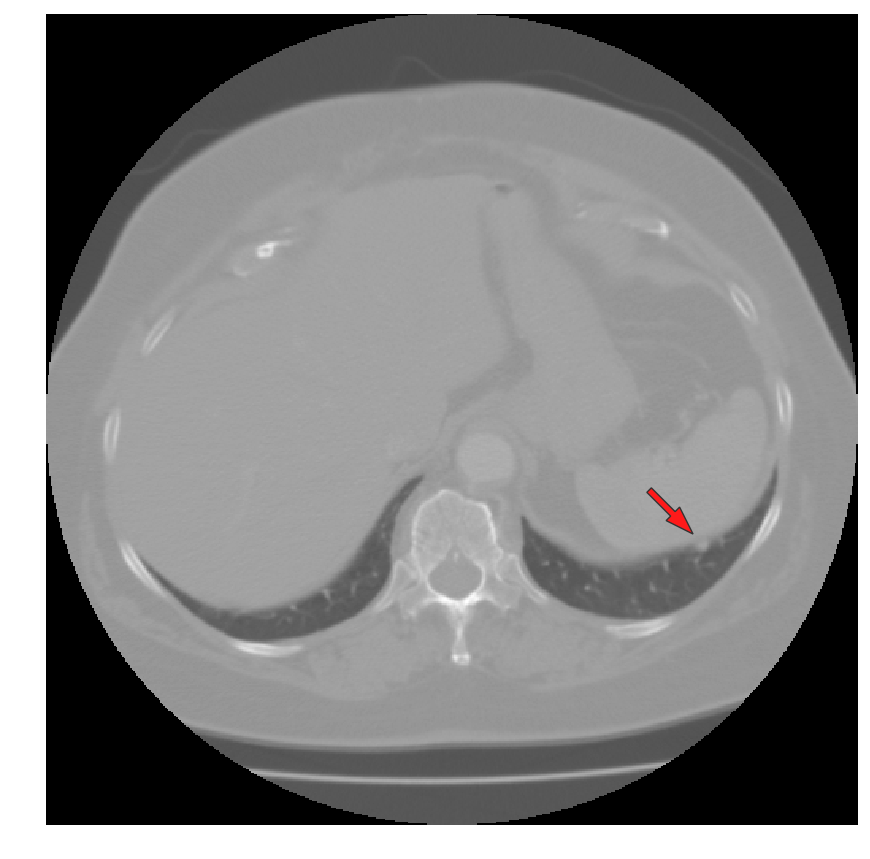}
		\caption{The nodule (at the tip of the red arrow) is shown in the context of the entire slice of the CT scan to which it belongs.}
	\end{subfigure}\hfill%
	\begin{subfigure}[t]{0.45\textwidth}
		\centering
		\includegraphics[width=\textwidth]{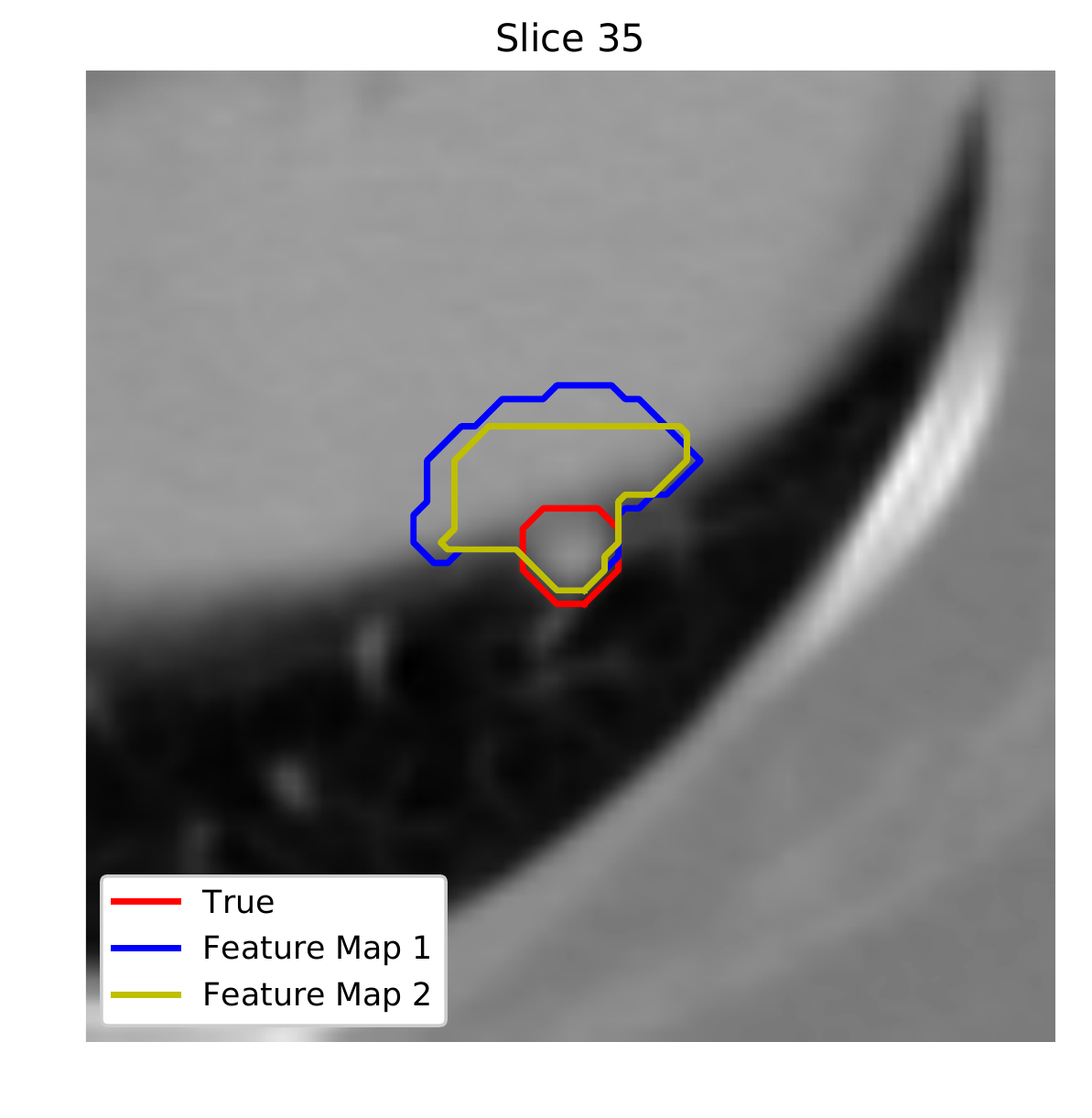}
		\caption{The center (35th) slice through the lung nodule image volume is shown with the true segmentation in red, along with final segmentation iteration of the LSML method under Feature Maps~1~and~2, shown in blue and yellow, respectively.}
	\end{subfigure}
	\caption{The lung nodule with minimal overlap score for both Feature Maps~1~and~2.}
	\label{fig:bad_nodule}
\end{figure}

Two outliers are observed (see Figure~\ref{fig:lidc_boxplots}) under both feature maps, with the lowest overlap scores of 0.0582 and 0.0732, for Feature Maps~1~and~2, respectively, which resulted from a poorly initialized segmentation of a juxta-pleural lung nodule in a region near the bottom of the lung, with the nearby lung wall and organs having image intensity values very close to those of the nodule, as can be seen in Figure~\ref{fig:bad_nodule}. Both of the LSML segmentations are poor because the final segmentation boundary of this nodule have relatively large section inside the lung wall, away from the nodule's boundary as annotated by the radiologists. This case is an outlier, but there are many juxta-pleural nodule cases (as well as other nodule anatomical location and density categories) where substantial improvements are observed from Feature Map~1 to Feature Map~2.

\begin{figure}[htbp]
	\centering
	\includegraphics[width=\textwidth,height=\textheight,keepaspectratio]{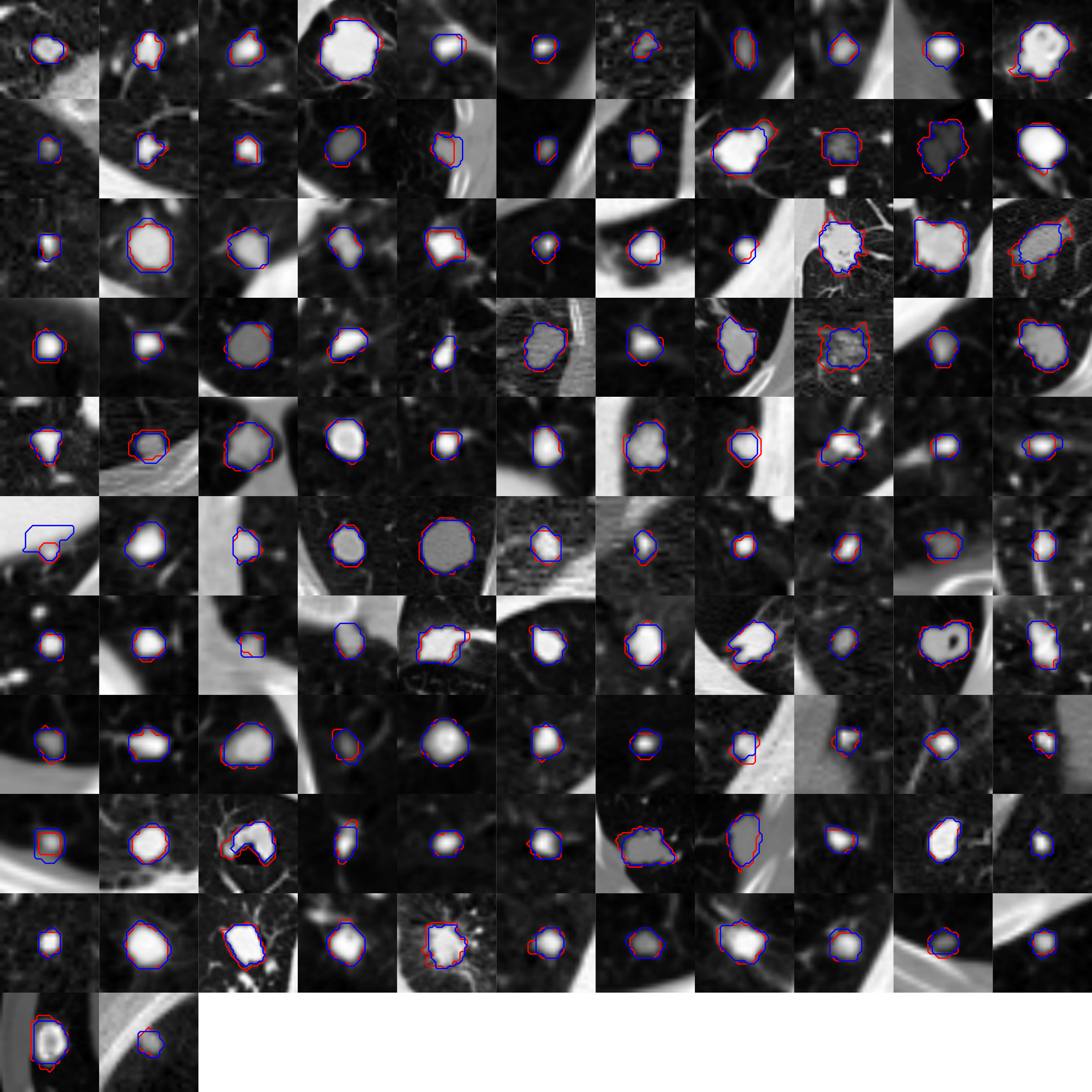}
	\caption{The center (35th) slice of each of the 112 lung nodules from the testing dataset. The red contour is produced by slicing the ground-truth segmentation through its center slice, and the blue contour is the approximate segmentation obtained by taking the center slice through the zero level surface produced by the LSML method using the superior Feature Map~2.}
	\label{fig:lidc_all}
\end{figure}

In Figure~\ref{fig:lidc_all}, we show the center (35th) slice through the image volume of each of the 112 lung nodules in the testing dataset, where the red curve represents the contour given by the slice through the ground-truth segmentation surface, and the blue curve represents the contour given by the slice through the approximate segmentation surface given by the LSML method using Feature Map~2. LSML performs well in a variety of contexts, including many juxta-pleural nodules (e.g., in row five, column two; and, row three, column eight), nodules with cavities (e.g., in row one, column eleven; and, row seven, column ten), non-solid nodules (e.g., in row four, column nine), irregularly-shaped nodules (e.g., in row nine, column three), spiculated nodules (e.g., in row ten, column five; and, row three, column nine), as well as the other nodule types shown in Figure~\ref{fig:lidc_all}. These visual results, together with the quantitative results discussed previously, demonstrate the effectiveness of the LSML method when applied to the lung nodule image segmentation problem in CT image volumes.

\subsection{Nodule Location and Density Type Segmentation Performance Analysis}
\label{subsection:results_by_type}
\begin{table}[htb]
	\centering
	\caption{Average overlap scores over the testing dataset organized by lung nodule anatomical location types for Feature Map~1 and Feature Map~2.}
	\label{tab:lsml_lidc_by_type_loc}
	\begin{tabular}{lll}
		\toprule
		Nodule Location Type & Feature Map 1 & Feature Map 2 \\
		\midrule
		($n=50$) Isolated       & 0.7124~($\pm$0.0900) & 0.7294~($\pm$0.0968) \\
		($n=29$) Juxta-pleural  & 0.6484~($\pm$0.1602) & 0.6742~($\pm$0.1504) \\
		($n=21$) Juxta-vascular & 0.7036~($\pm$0.0754) & 0.7346~($\pm$0.0794) \\
		($n=12$) Pleural-tail   & 0.7208~($\pm$0.0745) & 0.7517~($\pm$0.0652) \\
		\bottomrule
	\end{tabular}
\end{table}
We developed a graphical user interface (GUI) for categorizing nodules based on their anatomical location (isolated, juxta-pleural, juxta-vascular, and pleural-tail) and density appearance (sharp, semi-sharp, diffuse, and GGO). We assigned lung nodules in the testing dataset to these categories using this GUI and analyzed the corresponding segmentation performance of the LSML method. In Tables~\ref{tab:lsml_lidc_by_type_loc}~and~\ref{tab:lsml_lidc_by_type_img}, we present the average overlap scores computed over the testing dataset, for the various lung nodule anatomical location types and image density types, respectively. For each of the subcategories within both the location and density categories, there is an improvement in Feature Map~2 over Feature Map~1. Interestingly, the average overlap for pleural-tail nodules is greater than the average overlap score for isolated nodules for both feature maps, as shown in Table~\ref{tab:lsml_lidc_by_type_loc}; however, we note that the pleural-tail nodule category is less represented in the sample ($n=12$) compared to the isolated category ($n=50$), thus indicating the possibility of a statistical artifact for the pleural-tail category of nodules in this dataset.

\begin{table}[htb]
	\centering
	\caption{Average overlap scores for the testing dataset, organized by lung nodule density types for Feature Map~1 and Feature Map~2.}
	\label{tab:lsml_lidc_by_type_img}
	\begin{tabular}{lll}
		\toprule
		Nodule Density Type & Feature Map~1 & Feature Map~2 \\
		\midrule
		($n=25$) Sharp      & 0.7372~($\pm$0.1018) & 0.7412~($\pm$0.1113) \\
		($n=62$) Semi-sharp & 0.6874~($\pm$0.1162) & 0.7167~($\pm$0.1182) \\
		($n=23$) Diffuse    & 0.6759~($\pm$0.1062) & 0.7060~($\pm$0.0876) \\
		($n=\;\,2$) GGO     & 0.6270~($\pm$0.0255) & 0.6343~($\pm$0.0557) \\
		\bottomrule
	\end{tabular}
\end{table}

From Table~\ref{tab:lsml_lidc_by_type_loc}, we also observe that for both feature maps, the respective ordering of performances by lung nodule anatomical location type are identical, i.e., the best performance is observed for the pleural-tail types, followed by the isolated and juxta-vascular types, with the worst performance being the juxta-pleural types. The relatively poor performance on the juxta-pleural cases is expected because of the lack of image contrast along the lung nodule sides that are adjacent to the lung wall. The relative increases in accuracy from Feature Map~1 to Feature Map~2 are 2.39\%, 3.99\%, 4.41\%, 4.29\% for the isolated, juxta-pleural, juxta-vascular, and pleural-tail lung nodule location types, respectively. The larger relative increase is observed for lung nodules situated in more complex surroundings (inherent in juxta-pleural and pleural-tail nodules), which indicates that the ``glocal" features included in Feature Map~2 are helpful in these contexts.

Moving to the performance of the segmentation algorithm by image density category, given in Table~\ref{tab:lsml_lidc_by_type_img}, we notice once again that the relative ordering in performance by density subcategory between Feature Maps~1~and~2 is the same, i.e., the best performance is observed for sharp boundary types, followed by semi-sharp and diffuse types, with the worst performance occurring for the GGO (Ground Glass Opacity) density types. This performance ordering is not unexpected, since the results follow the expected ordering in complexity of nodule density, either internally or at the nodule boundary. We note, however, that there are only two GGO examples, and so the results for this nodule density subcategory are of course not statistically robust. The largest relative increase from Feature Map~1 to Feature Map~2 is 4.45\% for the diffuse lung nodule boundary density type, and the smallest relative increase is 0.54\% for the sharp density category, as one would expect. A 4.26\% relative increase from Feature Map~1 to Feature Map~2 is observed for the semi-sharp category. These findings indicate that the additional ``glocal" features added to Feature Map~1 to form Feature Map~2 (which provide additional neighborhood shape and image context) assist with the segmentation of the lung nodules in density subcategories that have the more complex boundaries.

\subsection{Feature Importance Analysis}
In this section, we consider only Feature Map~2 (as given in Table~\ref{tab:fmap2}), and we utilize the computed Random Forest feature importance (RFFI) metrics~\cite{friedman_elements_2001} to analyze the relative importance of the features used in the regression models, i.e., for $V^n$ in Eq.~\eqref{eq:lsapprox}. The RFFI metric is computed by randomly permuting the values for the $j$th feature on the out-of-sample training data (i.e., the training data not used in the bootstrap sampling procedure of the Random Forest algorithm), recording the increase in error due to the permutations, and averaging the error increase over the out-of-sample data. Larger RFFI values correspond to features whose value-permutations cause larger increases in error, on average, and thus a larger RFFI value signifies a more informative, and perhaps, more useful feature for segmentation. The RFFI metrics are normalized so that the sum of the computed feature importance values over all the features is equal to one.

\begin{table}
	\centering
	\caption{Feature importances by image and shape, and global and local types. Note that these feature sets are disjoint, and in total sum to one.}
	\label{tab:feat_imp_total}
	\begin{tabular}{lrr}
		\toprule
		{} &     Image &     Shape \\
		\midrule
		Global &  0.0629 &  0.0323 \\
		Local  &  0.7405 &  0.1643 \\
		\bottomrule
	\end{tabular}
\end{table}

In Table~\ref{tab:feat_imp_total}, we give the total feature importance (i.e., averaged over all iterations) for local, global, image, and shape feature types. From these results, it is clear that the local features are relatively more important (under the RFFI metric) than global features, and that the image features are relatively more important than shape feature types. 

In Table~\ref{tab:feat_imp_top_5}, we provide the top five features of each type, i.e., global image, global shape, local image, and local shape. Keeping in mind Figure~\ref{fig:sample-features}, we first describe the notation used. We denote by the acronym COM, the center-of-mass of the segmentation, which is computed using the first segmentation moments (see Appendix~\ref{subsection:lsml_features_global_shape}). The ``Normal ray" and ``COM ray" features are suffixed with a number of the form $\pm p$. A plus symbol ``+" indicates that an image sample is taken along the respective ray in the direction inward from the surface, whereas a minus symbol ``-" indicates sampling occurs in the outward direction. The respective ray has a length that is equal to the distance from the local coordinate $(i,j,k)$ to the center-of-mass. The number $p$ indicates the position along the ray as a percentage of the total length, starting from the initial position $(i,j,k)$. For example, ``COM ray +20\%" indicates an image sample taken at 20\% along the COM Ray from a given point $(i,j,k)$ inward toward the center of mass, whereas ``Normal ray +20\%" does the same, except that the sample is taken along the normal direction. We also indicate the image scale of the image features by $\sigma=0$ or $\sigma=3$, which indicates that the image volume was convolved with a Gaussian kernel with parameter $\sigma$, prior to computing the feature. Note that $\sigma=0$ means that no Gaussian-smoothing was applied.

\begin{table}[!h]
	\centering
	\caption{Top five features (as determined by the Random Forest feature importance metric) for each feature type. Definitions are given in Appendix~\ref{section:lsml_features}.}
	\label{tab:feat_imp_top_5}
	\fontsize{8}{10}\selectfont
	\begin{tabular}{l||llll}
		\toprule
		&Global Image & Global Shape & Local Image & Local Shape\\
		\midrule
		1. & Global image mean ($\sigma=0$)      & Isoperimetric ratio     & Normal ray +20\% ($\sigma=0$) & Slice area (y)\\
		2. & Global image edge ($\sigma=0$)      & Maximum distance to COM & COM ray +20\% ($\sigma=0$)    & Distance to COM\\
		3. & Global image std. dev. ($\sigma=0$) & Moment 2 (z)            & Normal ray +50\% ($\sigma=0$) & Slice area (z)\\
		4. & Global image std. dev. ($\sigma=3$) & Moment 1 (z)            & Local image edge ($\sigma=0$) & Slice area (x)\\
		5. & Global image mean ($\sigma=3$)      & Moment 1 (x)            & COM ray -100\% ($\sigma=3$)   & Slice area diff. (z)\\
		\bottomrule
	\end{tabular}
\end{table}

As shown in Table~\ref{tab:feat_imp_top_5}, the ``global image mean" feature (i.e., the average CT intensity value within the interior of the segmentation boundary) ranks highest amongst the global image features, with the ``global image edge" feature (i.e., the average image gradient-magnitude over the segmentation surface) ranking second highest. These features are natural since we expect lung nodules (and their corresponding CT image representation) in many cases to be homogeneous in the interior, as well as have homogeneous edge properties when the nodule margin is well-defined. The isoperimetric ratio is the feature that ranks highest amongst the global shape features, which is somewhat expected since the isoperimetric ratio measures the degree of sphericity of the segmentation surface of lung nodules, which are generally spherical with distortions. The second and third highest ranking global shape features are the maximum distance from the level surface to the center-of-mass and the second segmentation moment, ``Moment 2 (z)" along the depth direction in the volume, respectively. The ``maximum distance to the center-of-mass" feature is intuitively meaningful because it can be compared to the local distance feature at a given voxel coordinate and thus can have a stabilizing effect on the growth of the evolving segmentation boundary. The second moment along the depth direction measures how elongated the segmentation body is in the between-slice z-direction. The five highest ranking local image features include four ``Normal ray" and ``COM ray" image features, as well the local image edge feature. The top two local image features, ``Normal ray +20\%" and ``COM ray +20\%", provide image context slightly inward to a given voxel. The third and fifth most important local image features, ``Normal ray +50\%" and ``COM ray -100\%", respectively provide image context further inside and outside of the segmentation. It is intuitive that the latter would be helpful for making the decision of whether to grow or shrink the boundary in juxta-pleural cases since image information at the border alone is insufficient. The highest ranking local shape features include the ``slice area" features along each of the three coordinate axes, as well as the ``distance to center of mass" feature. Interestingly, the ``area slice difference" feature ranking highest among the possible three axes is along the depth (z) direction, which agrees with the importance of the global second segmentation moment feature, indicating that the superior and inferior anatomical directions are distinguished, perhaps for biological or medical reasons that we will not speculate on. All the features used are generic image and shape features that probe the local and global variations of each interim nodule segmentation determined by the algorithm, and the features work synergistically and effectively to obtain the final nodule segmentation.

\begin{figure}
	\centering
	\includegraphics[width=0.7\textwidth]{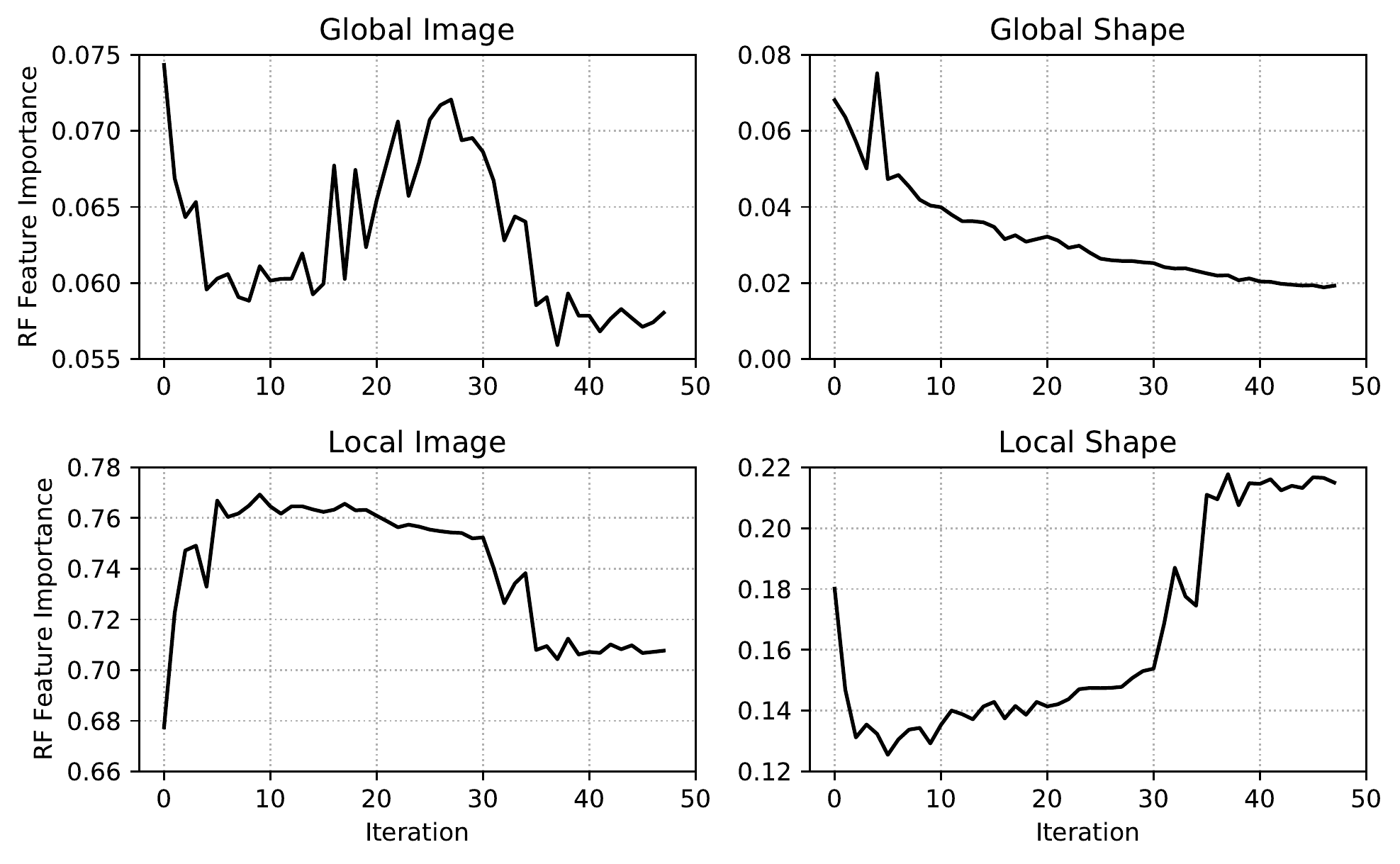}
	\caption{Feature importances by iteration number for global image, global shape, local image, and local shape feature types from top to bottom, left to right, respectively. Note that the vertical axis scales varies from plot to plot.}
	\label{fig:feat_imp_by_iter}
\end{figure}

In Figure~\ref{fig:feat_imp_by_iter}, we graph the RFFI metric against the iteration number for global and local, and shape and image, features. Note that the scale on the vertical axes varies from plot to plot due to the absolute differences by feature type, as was previously shown and discussed in Table~\ref{tab:feat_imp_total}. Here, we are interested in how the feature importances trend with the iteration number in the algorithm, i.e., over the course of the segmentation evolution in LSML. As shown in the bottom-left plot in Figure~\ref{fig:feat_imp_by_iter}, the local image features are relatively more important during approximately the first 30 iterations in the segmentation process. We also note that the global image features (see Figure~\ref{fig:feat_imp_by_iter}, top-left) are relatively more important initially during the first few iterations, and afterward, during iterations 20 through 30, approximately. In the top-right of Figure~\ref{fig:feat_imp_by_iter}, the global shape feature importance is seen largely to decline as the segmentation evolution progresses, indicating that the global shape features used in Feature Map~2 are more useful early in the segmentation process. This observation is reasonable, given that the global shape features used are very simple shape descriptors (e.g., volume or surface area, analogous to low order terms in a basis expansion sense). Such features would be expected to vary only slightly in the later iterations of the process as the segmentation boundary reaches a stable configuration. Lastly, we note that the local shape features (see Figure~\ref{fig:feat_imp_by_iter}, bottom-right) are relatively more important after about iteration 35 onward, which act in a complimentary way to the behavior observed in the local image features.

It is interesting to compare the feature importance curves in Figure~\ref{fig:feat_imp_by_iter} with the overlap score curves in Figure~\ref{fig:lidc_overlap_iter}. For the first twenty iterations, the overlap score increases more rapidly, on average, than for the subsequent iterations, and by the 30th-40th iterations, the average overlap score begins to asymptote. Thus, after about 35 iterations, the effect of all the features on the segmentation process is considerably diminished, since the average overlap score is concomitantly asymptoting to its final value (by the 47th iteration); and, only relatively small increases in the overlap score are observed during the final iterations (see Figure~\ref{fig:lidc_overlap_iter}, the dashed curve for Feature Map~2). In conjunction with Figure~\ref{fig:feat_imp_by_iter}, this suggests that the local shape features primarily play a role in making minor refinements during the last stages of the segmentation process. The image and shape feature extraction performed by the algorithm is thus able to make both large and small changes necessary for effective segmentation of the wide variety of lung nodule geometries and appearances in the LIDC dataset (see Figure~\ref{fig:lidc_all}), as observed from the comparative segmentation results in Table~\ref{tab:related_works_with_jaccard}.

\section{Conclusions}
\label{section:conclusions}
Lung nodule segmentation is a core component of the lung CAD pipeline, and accurate nodule segmentation poses unique challenges because of the different nodule varieties, their diversity, and location within the lung. The LSML method, a natural and direct machine learning extension of the level set image segmentation method, achieves an average Jaccard overlap of 0.7185~($\pm$0.1127) and Dice coefficient of 0.8362~($\pm$0.2178), which is comparable to the current state-of-the-art for lung nodule image segmentation. The mortality rate of lung cancer is large~\cite{cancer_mortality}, and lung CAD methods, if robustly validated in clinical settings, carry the potential to aid physicians towards increasing the survival rate of those afflicted. Accurate lung nodule segmentation is necessary for achieving this goal. The LSML method represents a general segmentation approach that can be employed and adapted in other applications, which may require domain-specific feature extraction operators.

\appendix

\subsection*{Disclosures}
No conflicts of interest, financial or otherwise, are declared by the authors.

\subsection*{Code Availability}
In the process of creating and experimenting with the LSML method, we have developed and implemented a Python software package which is freely available under BSD license. The code, which includes example usage on synthetic imagery, is available on GitHub:

\begin{itemize}
	\item {\linkable{https://github.com/notmatthancock/level-set-machine-learning/}}
\end{itemize}



\bibliography{refs}   
\bibliographystyle{spiejour}   


\vspace{2ex}\noindent\textbf{Matthew C. Hancock} is a scientific software developer at Enthought in Austin, Texas. He received a Ph.D. in applied mathematics from Florida State University, where research for this work was conducted.

\vspace{2ex}\noindent\textbf{Jerry F. Magnan} is an associate professor in the Mathematics Department at Florida State University, in the area of Applied and Computational Mathematics. He received a Ph.D. in physics from the University of Miami, and was a visiting assistant professor at Northwestern University, in the Department of Engineering Sciences and Applied Mathematics. His research interests involve the mathematical modeling and analysis of nonlinear phenomena in STEM systems, and the use of machine learning in scientific and industrial problems.

\section{Segmentation feature definitions}
\label{section:lsml_features}
Note that although we specify the features below assuming three-dimensional scalar fields for both the image and level set volumes, most of the features have natural one- and two-dimensional analogs. Below, we use $H$ to be the unit step function.

\subsection{Global shape features}
\label{subsection:lsml_features_global_shape}

A variety of geometrical features based purely on the current segmentation (i.e., the region where $u_{ijk}$ is positive) are utilized.

\begin{itemize}
	\item {\bf Boundary length}. This feature computes the surface-area of the zero level surface by approximating the integral, $\int_{\R^3} \|DH(u(x,y)\| \; dV$:
	\begin{equation}
	\label{eq:feat_global_length}
	L(u) = \sum_{ijk} (\|DH(u)\|)_{ijk}
	\end{equation}
	Central finite differences are used to approximate the gradient operator.
	
	\item {\bf Volume}. This feature approximates the area enclosed by the discretized level set by counting the number of coordinates where $u_{ijk}$ is positive:
	\begin{equation}
	\label{eq:feat_global_area}
	V(u) = \sum_{ijk} H(u_{ijk})
	\end{equation}
	
	\item {\bf Isoperimetric Ratio}. The isoperimetric ratio is defined as:
	\begin{equation}
	\label{eq:feat_global_iso}
	Q(u) = \frac{36\pi \cdot V(u)^2}{L(u)^3}
	\end{equation}
	This yields a value between zero and one, and is one when the boundary is volume (e.g., see reference~\cite{do_carmo_differential_nodate}). Thus, this gives a measure of sphericity, but also has constant value for other geometric planar shapes. This ratio becomes small when the surface area becomes unwieldy compared to its enclosed volume, which is generally an undesirable trait. Note that this feature is also defined for arbitrary dimensions, where in two-dimensions it gives a measure of circularity, for example.
	
	\item {\bf $p$th segmentation moments}. These comprise three features (along the $i,j,k$ axes). We often compute these features for $p=1,2$. When $p=1$, these yield the ``center of mass" of the region where $u_{ijk}$ is positive. When $p=2$, these features quantify the ``spread" of the region where $u_{ijk}$ is positive.
	\begin{equation}
	\label{eq:feat_global_moments}
	\bar{i}_p(u) = \frac{\sum_{ijk} i^p \cdot H(u_{ijk})}{V(u)}, \;\; \bar{j}_p(u) = \frac{\sum_{ijk} j^p \cdot H(u_{ijk})}{V(u)}, \;\; \bar{k}_p(u) = \frac{\sum_{ijk} k^p \cdot H(u_{ijk})}{V(u)}
	\end{equation}
	
	\item {\bf Mean, standard deviation, and maximum distance to center of mass}. The distance to center of mass feature (see Equation~\ref{eq:feat_local_dist_from_com}) is computed for coordinates near the zero level surface of $u$, and the average, standard deviation, and maximum over the computed distances are computed.
\end{itemize}

\subsection{Global image features}
\label{subsection:lsml_features_global_image}

Many global image features are computed by computing various statistics on the smoothed image values restricted to the region where $u_{ijk}$ is positive. Global image features can also be derived by restricting to image values near the boundary of the zero level set of $u_{ijk}$.

\begin{itemize}
	\item {\bf Mean inside}. This feature yields the average value of the image where $u_{ijk}$ is positive:
	\begin{equation}
	\label{eq:feat_global_image_mean}
	\bar{M}(u, M) = \frac{\sum_{ijk} (G_\sigma * M)_{ijk} \cdot H(u_{ijk})}{V(u)}
	\end{equation}
	
	\item {\bf Standard deviation inside}. This feature computes the variability of the image values inside the region where $u_{ijk}$ is positive:
	\begin{equation}
	\label{eq:feat_global_image_std}
	\bar{\sigma}_M(u, M) = \sqrt{\frac{\sum_{ijk} \left[(G_\sigma * M)_{ijk} - \bar{M} \right]^2 \cdot H(u_{ijk})}{V(u)}}
	\end{equation}
	
	\item {\bf Average edge strength}. This feature computes the total edge strength over the zero level surface of $u$, normalized by the surface area of the zero level surface. It is computed by approximating the surface integral, $\oint_{\{u = 0\}} \|D(G_\sigma * M)\| \; dS = \int_{\R^3} \|D(G_\sigma * M)\| \; \|D(H(u))\| \; dV$ and normalizing:
	\begin{equation}
	\label{eq:feat_global_edge}
	E_g(u, M) = \frac{1}{L(u)} \sum_{ijk} (\|D(G_\sigma * M)\|)_{ijk} \cdot (\|DH(u)\|)_{ijk}
	\end{equation}
	
\end{itemize}

\subsection{Local shape features}
\label{subsection:lsml_features_local_shape}

\begin{itemize}
	\item {\bf Distance from segmentation center of mass}. This feature could be considered both local and global, since it defined in terms of a previously computed global shape feature. This feature measures the distance from the coordinate, $(i,j,k)$, to current center of mass of the region where $u_{ijk}$ is positive, (as computed by the $p=1$ segmentation moments from Equation~\eqref{eq:feat_global_moments}).
	\begin{equation}
	\label{eq:feat_local_dist_from_com}
	D_m(i,j,k,u) = \sqrt{(i-\bar{i}_{p=1}(u))^2 + (j-\bar{j}_{p=1}(u))^2 + (k-\bar{k}_{p=1}(u))^2}
	\end{equation}
	
	\item {\bf Slice area}. This feature is semi-local and can be computed along each axes. The area of the slice corresponding to a given axes is computed.
	\begin{equation}
	\label{eq:feat_local_slice_area}
	A_x(i,j,k,u) = \sum_{jk} H(u_{ijk}), \;\; A_y(i,j,k,u) = \sum_{ik} H(u_{ijk}), \;\; A_z(i,j,k,u) = \sum_{ij} H(u_{ijk})
	\end{equation}
	
	\item {\bf Slice area absolute change}. This feature approximates the absolute value of the derivative of the previous slice area feature along a given axes by using a centered difference approximation.
	\begin{equation}
	\label{eq:feat_local_slice_area_change}
	DA_x(i,j,k,u) = \frac{1}{2} \bigl|A_x(i+1,j,k,u) - A_x(i-1,j,k,u)\bigr|
	\end{equation}
	The features along the other two axes (i.e., $DA_y$ and $DA_z$) are computed analogously.
\end{itemize}

\subsection{Local image features}
\label{subsection:lsml_features_local_image}

\begin{itemize}
	\item {\bf Image value}. This feature yields the value of the Gaussian-smoothed smoothed image at the coordinate, $(i,j,k)$:
	\begin{equation}
	\label{eq:feat_local_image}
	M_\sigma(i,j,k) = (G_\sigma * M)_{ijk}
	\end{equation}
	
	\item {\bf Edge strength (local)}. This feature yields the edge strength of the Gaussian-smoothed smoothed image at the coordinate, $(i,j,k)$. Centered differences are used on interior points, and forward and backward finite differences are used at the boundary to approximate the gradient:
	\begin{equation}
	\label{eq:feat_local_edge}
	E_\ell(i,j,k,M) = (\|D(G_\sigma * M)\|)_{ijk}
	\end{equation}
	
	\item {\bf Samples along normal}. Consider a particular point $x_0 = (i,j,k)$ and the associated normal to the level set at that point, $N_0 = -Du(x_0) / \|Du(x_0)\|$. Then the image at scale $\sigma$ (i.e., $M_\sigma = G_\sigma * M$) can be sampled along the line associated with $x_0$ and $N$, i.e.,
	\begin{equation}
	\label{eq:feat_local_normal}
	S_N(s) = M_\sigma(x_0 + s \cdot N_0)
	\end{equation}
	We sample $S_N(s)$ in the inward and outward direction (i.e., for positive and negative values of $s$) a distance $D_m(i,j,k,u)$ (i.e., the distance from $x_0$ to the center of mass) for a total of twenty samples (ten in each direction). See Fig.~\ref{fig:sample-features} for a conceptual diagram of this feature.
	
	\item {\bf Samples along center-of-mass ray}. Consider a particular point $x_0 = (i,j,k)$ and the computed center of mass, say $p_0$. Now consider the unit vector, $r_0 = (x_0 - p_0) / \|x_0 - p_0\|$. Samples of the image at scale $\sigma$ are obtained in the same manner as the ``samples along normal" feature with $r_0$ in the role of $N_0$. See Fig.~\ref{fig:sample-features} for a conceptual diagram of this feature.
	\begin{equation}
	\label{eq:feat_local_ray}
	S_N(s) = M_\sigma(x_0 + s \cdot r_0)
	\end{equation}
	
\end{itemize}

\listoffigures
\listoftables

\end{spacing}
\end{document}